\documentclass[aps,pra,floatfix,twocolumn,superscriptaddress,nofootinbib]{revtex4-1}
\usepackage{times}
\usepackage{amsmath,amsfonts,amssymb,verbatim,float,mathtools}
\usepackage[english]{babel}
\usepackage{hyperref}
\usepackage{txfonts}
\usepackage{epsfig}
\hypersetup{
     colorlinks   = true,
     citecolor    = blue,
    linkcolor=blue,
    urlcolor=blue
}
\pdfoutput=1

\def\mgb{MgB$_2\ $}

\begin{document}
\title{Shape resonances and shell effects in thin-film multiband superconductors}
\author{Aurelio Romero-Berm\'udez }
\address{University of Cambridge, Cavendish Laboratory, JJ Thomson Avenue, Cambridge, CB3 0HE, United Knigdom}
\author{Antonio M. Garc\'{\i}a-Garc\'{\i}a}
\address{University of Cambridge, Cavendish Laboratory, JJ Thomson Avenue, Cambridge, CB3 0HE, United Knigdom}
\address{CFIF, Instituto Superior T\'ecnico, Universidade de Lisboa, Avenida Rovisco Pais, 1049-001 Lisboa, Portugal}
\begin{abstract}
We study analytically the evolution of superconductivity in clean quasi-two-dimensional multiband supercon- ductors as the film thickness enters the nanoscale region by mean-field and semiclassical techniques. Tunneling into the substrate and finite lateral size effects, which are important in experiments, are also considered in our model. As a result, it is possible to investigate the interplay between quantum coherence effects, such as shape resonances and shell effects, with the potential to enhance superconductivity, and the multiband structure and the coupling to the substrate that tend to suppress it. The case of magnesium diboride, which is the conventional superconductor with the highest critical temperature, is discussed in detail. Once the effect of the substrate is considered, we still observe quantum size effects such as the oscillation of the critical temperature with the thickness but without a significant enhancement of superconductivity. In thin films with a sufficiently longer superconducting coherence length, it is, however, possible to increase the critical temperature above the bulk limit by tuning the film thickness or lateral size.
\end{abstract}
\maketitle

\section{Introduction}
Advances in sample growth and a better experimental control have substantially reinvigorated research in low-dimensional superconductivity \cite{xue2004science,xue2010natphys,shih2009science,Bose2010,wolf2009,wolf2012}. Refined scanning tunneling microscope techniques have been recently employed \cite{Bose2010,sangita2011prb} to study superconductivity in single isolated nanograins and also measure its size. It has also become possible  \cite{wolf2009,wolf2012} to measure with unprecedented precision the size dependence of the capacitance in nanoscale superconducting islands.
Epitaxial growth of superconducting thin films by adding single atomic layers \cite{xue2004science,xue2010natphys,shih2009science}, together with scanning tunneling microscope techniques, have permitted one to track the evolution of superconductivity as the film approaches the two-dimensional limit. For Pb, it was found that, on average, the critical temperature ($T_c$) is a decreasing function of the thickness. Oscillations, below the bulk critical temperature, were observed for intermediate thicknesses \cite{xue2004science}.

These oscillations in $T_c$ have been predicted theoretically \cite{Thompson1963} in the limit of no coupling to the substrate. However, the maxima of the oscillating pattern, usually referred to as shape resonances, were expected to correspond to $T_c$ substantially higher than in the bulk limit. These shape resonances occur as a consequence of an enhancement of the spectral density at the Fermi energy for thicknesses for which a new quantum state becomes available in the well potential that describes the confinement in the dimension perpendicular to the thin film. It was later realized \cite{allen1975,paskin1976,shanenko2007prb,chen2006quantum} that more realistic boundary conditions, including the charge neutrality condition, suppress this enhancement.  
In contrast, recent studies of heterostructures and interfaces based on cuprates, \cite{bozovic2008nature} iron pnictides\cite{xue2012natmat} and ${\rm LaAlO}_3/{\rm SrTiO}_3$ \cite{triscone2007science} heterostructures have clearly shown that superconductivity can occur on a single atomic layer and that the critical temperature can be enhanced with respect to the bulk limit. 

From these results it is not yet clear whether it is possible to enhance superconductivity in thin films by simply tuning the thickness. The enhancement observed in interfaces and heterostructures based on cuprates or iron-based superconductors, which is of special interest due to its high critical temperature, is difficult to model theoretically as there is not yet a good understanding of these materials. In contrast, magnesium diboride (MgB$_2$), a two-band superconductor, is a more attractive choice as it has a relatively high critical temperature ($\sim39$ K) but still is a conventional superconductor \cite{Nagamatsu2001} for which many theoretical tools are available. In recent experiments, it has been possible to grow good quality \mgb films of thicknesses less than $10$nm.\cite{Shimakage2008,Zhang2010,Zhang2013} Despite these advances the experimental control and growth techniques in \mgb films are still not comparable to Pb and other metallic superconductors, but the gap is rapidly closing. 

It is, therefore, timely to develop a theoretical description of quantum size effects in multiband thin-film superconductors that can clarify whether superconductivity is enhanced in some region of parameters. Indeed, several papers \cite{Karol2006,bianconi2003, bianconi2004,bianconi2010,bianconi2010a,aggsacramento2011} have already studied size effects in thin-film multiband superconductors, but a definitive answer is still missing: Bianconi and co-workers \cite{bianconi2003,bianconi2004,bianconi2010,bianconi2010a} were the first to suggest, by combining qualitative arguments with numerical simulations, that shape resonances could enhance superconductivity in \mgb and others multiband superconductors. Shell effects in multi-band superconductors,\cite{aggsacramento2011} though suppressed with respect to the one-band case, are still capable of increasing the critical temperature with respect to the bulk limit. By contrast, a numerical analysis of \mgb thin films \cite{Karol2006} that included the charge neutrality condition at the surface, but did not address directly the role of the substrate or shell effects, showed no enhancement of superconductivity. 

Here we generalize the one-band  model of Thompson and Blatt \cite{Thompson1963} for infinite thin film to the multiband case, including finite lateral size effects and the coupling to the substrate. Explicit analytical results are obtained by combining mean-field and semi-classical techniques. Therefore, our model is capable of accounting for the interplay of shape resonances and shell effects that can enhance superconductivity and the multi-band structure and the substrate that tend to suppress these coherence effects. All of these ingredients are important in the description of realistic thin films with negligible disorder. 

The main results of the paper are summarized as follows: for an infinite, free-standing, multiband thin film, we observe that the critical temperature is a non monotonous function of the thickness with maxima well above the bulk limit but smaller than in the one-band case. Once the substrate is included, the oscillations in the $T_c$ are significantly reduced. For \mgb, we do not observe a substantial enhancement of the critical temperature. For materials, such as metallic superconductors, with a longer coherence length, or weaker electron-phonon coupling, size effects are stronger and an enhancement of $T_c$ by tuning the thickness is feasible. A finite lateral size does also affect the average value of the shape resonances and induces shell effects with the potential to further increase $T_c$ with respect to the bulk limit.

The paper is organized as follows. First we review the one-band thin-film model of Thompson and Blatt. \cite{Thompson1963} Then, within a mean-field approach, we generalize it to the case of two-band superconductors including, by semiclassical techniques, the effect of a finite lateral size and the leakage of probability due to the coupling with the substrate. In the second part of the paper we explore the evolution of $T_c$ with thickness as a function of the lateral size, band structure parameters, electron-phonon interaction strength and the coupling to the substrate. We discuss the optimal settings to enhance superconductivity in realistic multiband thin films. Explicit results are presented for \mgb as well as for other band structure parameters and electron-phonon coupling constants.


\section{Background: One-band superconducting thin film}
We start with a brief summary of the Thompson and Blatt \cite{Thompson1963} mean-field description of shape resonances in free-standing--Dirichlet boundary conditions--one-band thin films. For a thin film of infinite lateral size, the one-particle electron eigenstates are simply 
\begin{equation}\label{wf1band}
\psi_{\vec k}(\vec r) \sim u_n(x)\frac{1}{L}e^{i(k_y y +k_z z)},
\end{equation}
where periodic boundary conditions have been imposed in the lateral dimensions, $y$ and $z$, and $\psi_{\vec k}(x=0)=\psi_{\vec k}(x=a)=0$ in the perpendicular dimension $x$ where $a$ is the thin-film thickness. The latter results in 
\begin{equation}\label{wf1bandb}
u_n(x)=\sqrt{\frac{2}{a}}\sin\left(\frac{n\pi x}{a}\right),\ n\in\mathbb{N}.
\end{equation}

For a finite-size system, where the spectrum is discrete, the BCS Hamiltonian in terms of a set of good quantum numbers, for instance $n$ in Eq. (\ref{wf1bandb}), is, 
\begin{equation}\label{|2}
\begin{split}
&H=\sum_{n,\sigma\alpha} \xi_{n\alpha}c^{\alpha^\dagger }_{n\sigma}c^\alpha_{n\sigma} + \hspace{-0.3cm}\sum_{n,n',\alpha,\beta}\hspace{-0.2cm} c^{\alpha^\dagger}_{n\uparrow}c^{\alpha^\dagger}_{n\downarrow}V_{\alpha n,\beta n'}c^\beta_{n'\downarrow}c^\beta_{n'\uparrow\beta}\\
&\hspace{0.5cm}V_{\alpha n,\beta n'}= -\lambda_{\alpha\beta} \tilde\delta_\alpha \mathcal{V} \int_\mathcal{V} \psi^2_{n\alpha}(\vec r) \psi^2_{n'\beta}(\vec r) d^3 \vec r
\end{split}
\end{equation}
where $V_{\alpha n,\beta n'}$ are the interaction matrix elements, $\lambda_{\alpha\beta}$  are the dimensionless inter- and intraband  coupling constants, $\mathcal{V}$ is the volume, $\tilde\delta_\alpha$ is the mean level spacing (the inverse of the density of states at the Fermi level), $\sigma$ is the spin index, $\alpha$ and $\beta$ are the band indices, and $\xi_{n\alpha}=\epsilon_{n\alpha}-\mu$ and $c_{n\sigma},\ c_{n\sigma}^\dagger$ are the usual quasiparticle annihilation and creation operators.

The maximum quantum number allowed, $n \equiv \nu$ in Eq. (\ref{wf1bandb}) must occur for a film thickness in the interval $[a_\nu,a_{\nu+1}]$, where \cite{Thompson1963}
\begin{equation}\label{anu}
a^3_\nu=\frac{\pi}{2N/V}\left(\frac{2}{3}\nu^3-\frac{\nu^2}{2}-\frac{\nu}{6}\right).
\end{equation}

For a thickness $a \in [a_\nu,a_{\nu+1}]$, the superconducting order parameter $\Delta$, obtained from the Hamiltonian given by Eq. (\ref{|2}) in the mean-field approximation, and chemical potential $\mu$ are given by \cite{Thompson1963}
\begin{equation}\label{gapTh}
\begin{split}
&\mu=\frac{\pi\hbar^2a}{\nu m}\left[\frac{N}{V}+\frac{\pi}{6a^3}\nu(\nu+\frac{1}{2})(\nu+1)\right]\ ,\\
&\Delta=\frac{\hbar\omega_D}{\mbox{sinh}[Ka/(\nu+1/2)]},\ K=\frac{1}{\lambda}\left(\frac{3N}{\pi V}\right)^{1/3},
\end{split}
\end{equation}
where $\lambda$ is the dimensionless coupling constant, $N/V$ is the number of electrons per unit volume and $\hbar \omega_D$ is the Debye energy.

\begin{figure}[H]
\center
\includegraphics[scale=0.85]{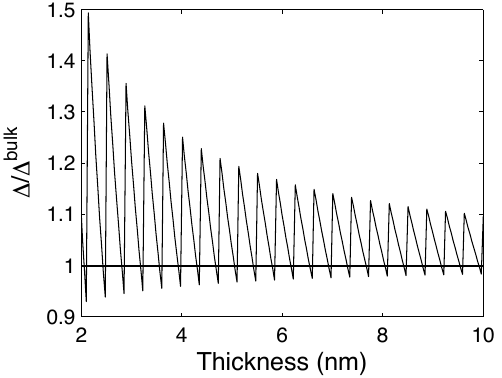}
\vspace{-0.4cm}
\caption{The superconducting order parameter $\Delta$ at $T=0$ K in units of the bulk value $\Delta^{bulk}$ as a function of the film thickness for a free standing one-band superconducting thin film. \cite{Thompson1963} Shape resonances are clearly observed as the thickness is increased each time a new state becomes available in the direction perpendicular to the film.}
\label{Blatt_gap}
\end{figure}

For sufficiently small thicknesses $a$, $\nu = 0$ and the system is purely two dimensional. However, as the thickness is increased, eventually $\nu=1$, which corresponds to a subband of allowed states in the perpendicular dimension. This increases the spectral density around the Fermi energy. The dimensionless electron-phonon coupling constant is proportional to the spectral density, so an enhancement of the latter increases the former. As a consequence, the order parameter and the critical temperature increase as well. This is what is usually called a shape resonance. As the thickness further increases, there exists a region in which still $\nu = 1$. The spectral density gradually becomes smaller and the critical temperature decreases. For the smallest thickness for which $\nu = 2$ a new sub-band is available which induces a new enhancement of superconductivity. As is depicted in Fig. \ref{Blatt_gap}, that results in a saw-like dependence of the superconducting gap and the critical temperature as a function of the film thickness.
\vspace{-0.2cm}

\section{Two-band superconducting thin film}
\subsection{Free-standing film model}\label{sec:free_model}
In this section, we extend the Blatt and Thompson formalism to the case of a two-band superconductor. Assuming again periodic boundary conditions in the lateral dimensions and Dirichlet boundary conditions in the perpendicular dimension, we have the equivalent of Eq. (\ref{wf1band}) in each band, but now with two quantum numbers: $n_\sigma,\ n_\pi$, analogous to $n$ in Eq. (\ref{wf1bandb}).
The dispersion relation is still quadratic,
\begin{equation}\label{bands}
\begin{split}
\hspace{-1mm}&\epsilon_{\vec k}^{(\sigma)}=\frac{\hbar^2}{2}\left[\frac{(k_y^{(\sigma)})^2}{m_{2\sigma}}+\frac{(k_z^{(\sigma)})^2}{m_{3\sigma}}\right]+\frac{\hbar^2}{2m_{1\sigma}}\left(\frac{n_\sigma\pi}{a}\right)^2\ ,\\
\hspace{-2mm}&\epsilon_{\vec k}^{(\pi)}=\frac{\hbar^2}{2}\left[\frac{(k_y^{(\pi)})^2}{m_{2\pi}}+\frac{(k_z^{(\pi)})^2}{m_{3\pi}}\right]+\frac{\hbar^2}{2m_{1\pi}}\left(\frac{n_\pi\pi}{a}\right)^2+e_{0\pi},
\raisetag{40pt}
\end{split}
\end{equation}
but with an offset $e_{0\pi}$ between the two bands.
A mean-field treatment of the microscopic Hamiltonian given by Eq. (\ref{|2}) for the two-band system \cite{bianconi2010,bianconi2010a,aggsacramento2011,Karol2006} results in the following two coupled gap equations  at zero temperature:

\begin{equation}\label{gap_def}
\begin{split}
&\Delta_\sigma=-\frac{1}{2}\sum_{k'}\left[\frac{\Delta_\sigma V_{\sigma k \sigma k'}}{\sqrt{(\epsilon_\sigma-\mu)^2+\Delta_\sigma ^2}}+\frac{\Delta_\pi V_{\sigma k \pi k'}}{\sqrt{(\epsilon_\pi-\mu)^2+\Delta_\pi ^2}}\right],\\
&\Delta_\pi=-\frac{1}{2}\sum_{k'}\left[\frac{\Delta_\sigma V_{\pi k \sigma k'}}{\sqrt{(\epsilon_\sigma-\mu)^2+\Delta_\sigma ^2}}+\frac{\Delta_\pi V_{\pi k \pi k'}}{\sqrt{(\epsilon_\pi-\mu)^2+\Delta_\pi ^2}}\right],
\end{split}
\end{equation}
while at finite temperature a factor $\tanh[\sqrt{(\epsilon_\alpha-\mu)^2+\Delta_\alpha^2}/2k_BT]$ multiplies each term on the right-hand side of the equations; the index $\alpha$ takes the value of the index of the order parameter in the corresponding term.

 $V_{\alpha k\beta k'}$ are the interaction matrix elements corresponding to two intraband coupling constants and two interband coupling constants,
 \vspace{-4mm}
\begin{equation}\label{mat_elem}
\begin{split}
V_{\alpha k\beta k'}&=-J_{\alpha\beta}\int_\mathcal{V} d^3\vec{r}|\psi_{\bf k}^{(\alpha)}(\vec r)|^2|\psi_{\bf k'}^{(\beta)}(\vec r)|^2 \\
&=-\frac{J_{\alpha\beta}}{a L^2}\left(1+\frac{1}{2}\delta_{n_{\alpha}n_{\beta}'}\right)\ ,
\end{split}
\end{equation}
where $\alpha$ and $\beta$ take the value of the band labels $\sigma$ and $\pi$. $J_{\alpha\beta}=\lambda_{\alpha\beta}\mathcal{V}\tilde\delta_\alpha$ and $\psi_{\bf k}^{(\alpha)}(\vec r)$  are of the form given by Eq. (\ref{wf1band}).

We then substitute Eq. (\ref{mat_elem}) into (\ref{gap_def}) and perform the sums in $k_y^{(\sigma)},\ k_z^{(\sigma)},\ k_y^{(\pi)},\ k_z^{(\pi)}$ by introducing the two-dimensional density of states in each band.\footnote{For the dispersion relations given by Eq. (\ref{bands}), the two-dimensional density of states is $g_{2D}^{(\alpha)}=\frac{L^2\sqrt{m_{1\alpha} m_{2\alpha}}}{\pi \hbar^2}$.} After carrying out the resulting integrations, we obtain the following system of two coupled equations at $T=0$:
\begin{equation}\label{gapT}
\begin{split}
&\Delta_\sigma=\frac{1}{2aL^2} \left[\Delta_\sigma J_{\sigma\sigma}g_{2D}^{(\sigma)} f(\sigma) + J_{\sigma\pi}\Delta_\pi g_{2D}^{(\pi)} f(\pi) \right]\ ,  \\
&\Delta_\pi=\frac{1}{2aL^2} \left[\Delta_\pi J_{\pi\pi}g_{2D}^{(\pi)} f(\pi) + J_{\pi\sigma}\Delta_\sigma g_{2D}^{(\sigma)} f(\sigma) \right]\ , 
\end{split}
\end{equation}
with $f(\alpha)=\left( \nu_\alpha +\frac{1}{2}\right) \text{asinh}\left(\frac{\hbar\omega_D}{\Delta_\alpha}\right)$.

For the calculation of the critical temperature a simple algebraic manipulation of Eq. (\ref{gap_def}) leads to the following relation between the two gaps,
\begin{equation}\label{TcT0}
\begin{split}
&\frac{1+\frac{1}{2}\sum_{k'}\frac{V_{\sigma k \sigma k'}}{\sqrt{(\epsilon_\sigma-\mu)^2+\Delta_\sigma^2}}\mbox{tanh}\frac{\sqrt{(\epsilon_\sigma-\mu)^2+\Delta_\sigma^2}}{2k_BT}}{\frac{1}{2}\sum_{k'}\frac{V_{\pi k \sigma k'}}{\sqrt{(\epsilon_\sigma-\mu)^2+\Delta_\sigma^2}}\mbox{tanh}\frac{\sqrt{(\epsilon_\sigma-\mu)^2+\Delta_\sigma^2}}{2k_BT}}=\\
&=\frac{\frac{1}{2}\sum_{k'}\frac{V_{\sigma k \pi k'}}{\sqrt{(\epsilon_\pi-\mu)^2+\Delta_\pi^2}}\mbox{tanh}\frac{\sqrt{(\epsilon_\pi-\mu)^2+\Delta_\pi^2}}{2k_BT}}{1+\frac{1}{2}\sum_{k'}\frac{V_{\pi k \pi k'}}{\sqrt{(\epsilon_\pi-\mu)^2+\Delta_\pi^2}}\mbox{tanh}\frac{\sqrt{(\epsilon_\pi-\mu)^2+\Delta_\pi^2}}{2k_BT}}.
\end{split}
\end{equation}

Repeating the steps described previously to obtain Eq. (\ref{gapT}) and taking the limits $\Delta_\sigma,\Delta_\pi\rightarrow 0$ and $T\rightarrow T_c$ gives,
\begin{equation}\label{TcT}
\frac{1-\frac{J_{\sigma,\sigma}g_{2D}^{\sigma}}{2aL^2}(\nu_\sigma+\frac{1}{2})F(T_c)}{-\frac{J_{\pi,\sigma}g_{2D}^{\sigma}}{2aL^2} (\nu_\sigma+\frac{1}{2})F(T_c)}=\frac{-\frac{J_{\sigma,\pi}g_{2D}^{\pi}}{2aL^2}(\nu_\pi+\frac{1}{2})F(T_c)}{1-\frac{J_{\pi,\pi}g_{2D}^{\pi}}{2aL^2} (\nu_\pi+\frac{1}{2})F(T_c)}.
\end{equation}

We have used that for $b\gg1$, $\int_0^b\text{d}x\tanh(x)/x\simeq\log(\frac{4e^\gamma}{\pi}b) = F(T)$, with $\gamma$ the Euler-Mascheroni constant and $b=\frac{\hbar\omega_D}{2k_BT_c}$. We note that $\nu_\pi$ and $\nu_\sigma$, i.e., the generalization of $\nu$ in Thompson and Blatt's one-band model given by Eqs. (\ref{anu}) and (\ref{gapTh}), are the maximum integers for which the condition $|\epsilon_{\vec k}^{(\alpha)}-\mu|\leq \hbar \omega_D$ holds. 
The superconducting gaps $\Delta_{\alpha}(T=0)$ and the critical temperature $T_c$ are therefore obtained by solving  Eqs. (\ref{gapT}) and (\ref{TcT}), respectively. Similarly, the chemical potential $\mu$ is obtained analytically from
\begin{equation}\label{NT}
\begin{split}
N&=\int_0^\mu [g_{3D}^{\sigma}(E)+g_{3D}^{\pi}(E)]\mbox{d}E=\\
&=\sum_{j=1}^{\nu_\sigma}Ê\int_0^{\mu-\eta_j^{\sigma}} g_{2D}^{\sigma}\mbox{d}\xi_{xy}^{\sigma}+\sum_{j=1}^{\nu_\pi} \int_0^{\mu-\eta_j^{\pi}} g_{2D}^{\pi}\mbox{d}\xi_{xy}^{\pi}=\\
&=\sum_{j=1}^{\nu_\sigma} g_{2D}^{\sigma}(\mu-\eta_j^{\sigma})+\sum_{j=1}^{\nu_\pi} g_{2D}^{\pi}(\mu-\eta_j^{\pi})\ ,
\end{split}
\end{equation}
where $\eta_j^{\pi}(a)=e_{0\pi}+\frac{\hbar^2\pi^2j^2}{2m_{1\pi}a^2}$ and $\eta_j^{\sigma}(a)=\frac{\hbar^2\pi^2j^2}{2m_{1\sigma}a^2}$. 

Using Faulhaber's formula for the second power sum of the first $n$ positive integers, it is also straightforward to obtain and explicit expression for the chemical potential,
\begin{equation}\label{muT}
\begin{split}
\mu&=\frac{a\pi\hbar^2}{\nu_\sigma m^{*\sigma}+ \nu_\pi m^{*\pi}}\left\{\frac{N}{V}+\frac{\pi }{2a^3}\left[\frac{m^{*\sigma}h(\nu_\sigma)}{m_{1\sigma}}+\frac{m^{*\pi}h(\nu_\pi)}{m_{1\pi}}\right]\right\}\\
&+\frac{e_{0\pi}\nu_\pi m^{*\pi}}{\nu_\sigma m^{*\sigma}+ \nu_\pi m^{*\pi}},
\end{split}
\end{equation}
with $m^{*\alpha}=\sqrt{m_{2\alpha}m_{3\alpha}}$ and $h(\nu_\alpha)=\frac{\nu_\alpha^3}{3}+\frac{\nu_\alpha^2}{2}+\frac{\nu_\alpha}{6}$. 

In order to find $a=a(\nu_\sigma,\nu_\pi)$ we first assume a value of $\nu_\sigma,\ \nu_\pi$ such that $\mu \simeq \eta_{\nu_\pi}^{\pi}(a)\simeq\eta_{\nu_\sigma}^{\sigma}(a)$, i.e.,
\begin{equation}
\nu_\sigma\simeq\sqrt{\frac{m_{1\sigma}}{m_{1\pi}}\nu_\pi^2+\frac{2m_{1\sigma}a^2}{\pi^2\hbar^2}e_{0\pi}}\ .
\end{equation}

Substituting, for every $\nu_\pi=1,2,3...$,  both $\mu$ and $\nu_\sigma$ in Eq. (\ref{muT}), we solve for $a$ and then calculate all of the possible states that are occupied as the thickness increases.
In order to proceed, we start with arbitrary values of $\nu_\sigma,\ \nu_\pi$ and assume that either $\eta_{\nu_\sigma}^{\sigma}>\eta_{\nu_\pi}^{\pi}$ or $\eta_{\nu_\sigma}^{\sigma}<\eta_{\nu_\pi}^{\pi}$, which results either in $\mu\simeq\eta_{\nu_\sigma}^{\sigma}$ or in $\mu\simeq \eta_{\nu_\pi}^{\pi}$ where, in order to simplify the notation, we have dropped the dependence in the thickness $a$. By substituting these expressions into Eq. (\ref{NT}) we obtain two equations for $a$ which are solved numerically. Once $a$ is obtained, we check which assumption ($\eta_{\nu_\sigma}^{\sigma}>\eta_{\nu_\pi}^{\pi}$ or $\eta_{\nu_\sigma}^{\sigma}<\eta_{\nu_\pi}^{\pi}$) holds and obtain the chemical potential from Eq. (\ref{muT}). \\
From these solutions we get the chemical potential given by Eq. (\ref{muT}), the gap given by Eq. (\ref{gapT}), and the critical temperature given by Eq. (\ref{TcT}) for a fixed $\nu_\sigma,\ \nu_\pi$ and $a\in[a_{\nu_\sigma\nu_\pi},a_{\tilde \nu_\sigma \tilde\nu_\pi}]$, where $\nu_\sigma\nu_\pi,\ \tilde \nu_\sigma \tilde\nu_\pi$ are consecutive states of the spectrum.

\subsection{Role of the substrate}\label{sec:role}
In realistic circumstances, a thin film is never isolated. It is usually placed on a substrate so there is some probability for the electrons to hop from the film into the substrate or at least penetrate a finite distance in it. Generally, this can be taken into account by assigning a finite lifetime to the quantized states and also by modeling the substrate thin-film interface by a potential more realistic than an infinite well.\\ 
The details of the coupling between the substrate and the thin film are very sensitive to the substrate material and the nature of the interface which depends on the growth techniques. A detailed microscopic description of the tunneling process is beyond the scope of this paper.\\
 Here we use recent experimental results \cite{Zhang2013} for \mgb and assume a linear dependence for the level broadening with the film thickness. We note that both the energy spectrum and the wavefunctions inside the film are modified by tunneling into the substrate. The latter has a direct impact on the interaction matrix elements given by Eq. (\ref{mat_elem}), while the former smoothes out the one dimensional density of states from a set of isolated Dirac's delta functions to a distribution with broader peaks. \\
In order to proceed, we write the density of states \cite{Brack} as,
\begin{equation}\label{DoS_expression}
g^\alpha(E)=\frac{dn^\alpha(E)}{dE}\left\{1+2\sum_{l=1}^\infty \kappa(l)\cos\left[2 l\pi n^\alpha(E) \right]\right\}\ ,
\end{equation}
where $n^\alpha(E)=\sqrt{(E-e_{0\alpha})/E^\alpha_0}$ and $n\in\mathbb{N}$ in the case of a infinite well potential, $E_0^\alpha=\hbar^2\pi^2/(2m_{1\alpha}a^2)$, $e_{0\sigma}=0$ and $e_{0\pi}\neq0$. For no tunneling into the substrate, $\kappa(l) = 1$ and we recover the usual expression in terms of Dirac delta functions. Tunneling or any other decoherence mechanism makes the system open which effectively induces level broadening, namely, the eigenvalues become complex. A natural way to mimic this effect is to introduce a cutoff,
\begin{equation}\label{weight_factor}
\kappa(l)\approx e^{-(lt/\tau)^2},
\end{equation}
where $t=2m_{1\alpha}a/\hbar k^\alpha_n$ and  $\tau$ is the typical lifetime of a quasiparticle at that energy. Physically, it is the typical time that an electron stays in the thin film. The specific functional form of $\kappa(l)$ depends to some extent on the mechanism that causes decoherence. The above result is obtained  (see Sec. $5.5$ in Ref. \cite{Brack} for more details) by replacing the original Dirac delta functions with Gaussians of width $\Gamma\sim\hbar/\tau$. 

Regarding the energy quantization, we model the thin film plus the substrate as a semi-infinite potential well, infinite in the film/vacuum interface and finite in the film/substrate interface. The height of the step corresponds to the  mismatch between the bulk Fermi levels of the film and substrate materials.
Furthermore it will also be assumed that the lifetime of all of the states is described by a single parameter since the total energy of the states is always very close to the Fermi level. 
\vspace{-0.5cm}
\subsection{Chemical potential of a two-band film on a substrate}
In order to compute the chemical potential in the presence of the substrate, we apply the Poisson summation formula, \cite{Apostol} given in Eq. (\ref{Poiss}), to Eq. (\ref{NT}) (see Appendix). This results in the following transcendental equation for $\mu$:
\begin{equation}\label{muD}
\begin{split}
N=&\sum_{j=1}^{\nu_\sigma} g_{2D}^{\sigma}(\mu-\eta_j^{\sigma})+\sum_{j=1}^{\nu_\pi} g_{2D}^{\pi}(\mu-\eta_j^{\pi})=\\
=&\sum_\alpha  g_{2D}^{\alpha}\left\{ \frac{2}{3\sqrt{E_0^\alpha}}(\mu-e_{0\alpha})^{3/2}-\frac{\mu-e_{0\alpha}}{2}+\right.\\
&\sum_{l=1}^\infty\left[-\frac{\sqrt{(\mu-e_{0\alpha}) E_0^\alpha}}{\pi^2 l^2}\cos\left(2\pi l \sqrt{\frac{\mu-e_{0\alpha}}{E_0^\alpha}}\right)\right.\\
&\left.\left.+\frac{E_0^\alpha}{2\pi^3 l^3}\sin\left(2\pi l \sqrt{\frac{\mu-e_{0\alpha}}{E_0^\alpha}}\right)\right]e^{-(lt/\tau_{\alpha})^2}\right\},
\end{split}
\end{equation}
where the sum over $\alpha$ refers to both bands, $E_0^\alpha=\hbar^2\pi^2/(2m_{1\alpha}a^2)$, $e_{0\sigma}=0$ and $e_{0\pi}\neq0$.
\subsection{Matrix elements and critical temperature of a two-band film on a substrate}\label{sec:mat_elem}
Before we proceed to the computation of the critical temperature we study the modification of the interaction matrix elements by the coupling to the substrate. 
We expect smaller matrix elements than those given by the infinite potential well model\cite{Thompson1963} since the amplitude of probability for all of the states inside the well is smaller. Moreover, since the energy states have a finite lifetime, the interaction is weighted by a smooth density of states, resulting in smooth shape resonances. 
The eigenstates inside a semi-infinite potential well are
 \begin{equation}\label{eigenstates}
 u_n^{(in)}(x)=A_n\sin(k_nx)\ ,
 \end{equation} 
 where $k_n$ is the solution of the quantization condition: $k_na=n\pi-2\arctan(-k/\tilde \kappa_n)$, $\tilde \kappa_n=\frac{m_{in}}{m_{out}}\kappa_n=\frac{m_{in}}{m_{out}}\sqrt{\frac{2m_{out}}{\hbar^2}(V_0-E_n)}$ obtained after imposing the BenDanield-Duke boundary conditions:\cite{BenDaniel1966}
\begin{equation}\label{bdy_cond}
\frac{1}{m_{out}}\left.\frac{\partial u_n^{(out)}}{\partial x}\right|_{x=b}=\frac{1}{m_{in}}\left.\frac{\partial u_n^{(in)}}{\partial x}\right|_{x=b}\ ,
\end{equation}
where
$b$ is the position of the interface and $m_{out}$ ($m_{in}$) is the effective mass outside (inside) of the well. We have taken the free electron mass for $m_{out}$ and $m_{1\alpha}$ for $m_{in}$.\\
The matrix elements resulting from the above expression for $u_n$ lead to a system of equations for two momentum-dependent superconducting order parameters, which are difficult to solve. In order to have a more tractable expression, we approximate the interaction of all of the states by the interaction of the states whose energies are equidistant between those corresponding to the highest and lowest occupied levels. If the highest (lowest) occupied states were used to estimate the interaction, the eigenstates' leakage out of the film would be  overestimated (underestimated). In our notation, this means the replacement of $A_n$ by $A_{m_\alpha}$ and $k_n$ by $k_{m_\alpha}$, where $m_\alpha$ refers, from now on, to the state whose energy is the closest to being equidistant from the highest- and the lowest-energy states. Moreover, we approximate $k_n$ in the argument of the sine of Eq. (\ref{eigenstates})  by $n \pi /a$, while leaving the amplitude $A_n$ unchanged.

With these further simplifications, the matrix elements are
\begin{equation}\label{mat_elem_K}
V_{\alpha k\beta k'}=-\frac{aJ_{\alpha\beta}}{4L^2}K_{\alpha\beta}\left(1+\frac{1}{2}\delta_{n_{\alpha}n_{\beta}'}\right)\ ,
\end{equation}
with $K_{\alpha\beta}=|A_{m_\alpha}|^2|A_{m_\beta}|^2$, given explicitly in Eq. (\ref{K_alpha_beta}). We now take into account the smoothed spectrum given by Eq. (\ref{DoS_expression}) due to the substrate. The sums of the matrix elements in Eq. (\ref{TcT0}) are simplified to
\begin{equation}\label{mat_elem_DoS}
\begin{split}
&\hspace{0.5cm}\sum_{n'_\beta=1}^{\nu_\beta}\left(1+\frac{1}{2}\delta_{n_{\alpha}n_{\beta}'}\right)= \frac{1}{2}+\int_{e_{0\beta}}^\mu g^\beta(E) dE=f(\beta)\ ,\\
&f(\beta)\equiv\frac{1}{2}+\sqrt{\frac{\mu-e_{0\beta}}{E_0^\beta}}+\sum_{l=1}^\infty \frac{e^{-\frac{tl}{\tau_\beta}}}{\pi l}\sin \left(2\pi l \sqrt{\frac{\mu-e_{0\beta}}{E_0^\beta}}\right)\ ,
\end{split}
\end{equation}
where $e_{0\sigma}=0$ and $e_{0\pi}\neq0$. Finally we substitute Eqs. (\ref{mat_elem_K}) and (\ref{mat_elem_DoS}) into Eq. (\ref{TcT0}) to obtain
\begin{equation}\label{TcD}
\begin{split}
&\frac{1-\frac{aJ_{\sigma,\sigma}g_{2D}^{\sigma}}{8L^2}K_{\sigma\sigma}f(\sigma)F(T_c)}{-\frac{aJ_{\pi,\sigma}g_{2D}^{\sigma}}{8L^2} K_{\pi\sigma} f(\sigma)F(T_c)}=\frac{-\frac{aJ_{\sigma,\pi}g_{2D}^{\pi}}{8L^2}K_{\sigma\pi}f(\pi)F(T_c)}{1-\frac{aJ_{\pi,\pi}g_{2D}^{\pi}}{8L^2} K_{\pi\pi}f(\pi)F(T_c)},
\end{split}
\end{equation}
where $F(T_c)=\log(\frac{4e^\gamma}{\pi}\frac{\hbar\omega_D}{2k_BT_c})$, $K_{\alpha\beta}$ is given in Eq. (\ref{K_alpha_beta}), and $f(\alpha)$ is given in Eq. (\ref{mat_elem_DoS}). The final step to compute the critical temperature is to solve Eq. (\ref{TcD}) for $T_c$ and different thicknesses. 

\subsection{Lateral size effects in a two-band superconducting thin film}\label{sec:finite_lateral_size}
We now study the case in which the thin-film lateral size dimensions ($y=L_1$ and $z=L_2$) become comparable to the film thickness $a$. 
We will not go through the details of the calculations regarding the modification of the two-dimensional density of states. A detailed derivation can be found in Ref. \cite{Garcia2011}. The underlying idea is to use the semiclassical approximation, valid in the limit $(k_FL)^{-1}\ll 1$ with $L$ in this case the lateral film size, to write down the density of states as a sum over the classical periodic orbits of the two-dimensional system. The density of states is an oscillatory function of the energy around the Fermi level so, in principle, it should enter explicitly in the sums over $k_y$ and $k_z$ which are needed to solve the gap equation (\ref{gap_def}).
However, it was demonstrated in Ref.\cite{Garcia2011} that the density of states can be taken out of the integral, provided it is smoothed out, as follows:
\begin{equation}\label{factors1}
\tilde g_{2D}^{(\alpha)}\simeq g_{2D}^{(\alpha)}[1+\overline{g}^{(\alpha)}+ g _l^{(\alpha)}]\ ,
\end{equation}
where the correction $\overline g^{(\alpha)}$ is an average term, while $g_l^{(\alpha)}$ is an oscillatory term that depends on the length $l$ of the periodic orbits in the $yz$ plane. These corrections are,
\begin{equation}\label{factors2}
\begin{split}
&\overline{g}^{(\alpha)}=-\frac{L_1+L_2}{  k_{yz}^{(\alpha)}L_1L_2},\\
& g _l^{(\alpha)}= g _{12}^{(2\alpha)}-\frac{1}{2}{g} _{1}^{(1\alpha)}-\frac{1}{2}{g} _{2}^{(1\alpha)},
\end{split}
\end{equation}
and
\begin{equation}\label{factors3}
\begin{split}
&{g} _{12}^{(2\alpha)}=\sum_{\vec n\neq\vec0}^{\infty} J_0(  k_{yz}^{(\alpha)}L_{\vec n}^{1,2})\times K_0(L_{\vec n}^{1,2}/\xi^{(\alpha)})\ , \\ 
&{g} _{1}^{(1\alpha)}=\frac{4}{  k_{yz}^{(\alpha)}L_2}\sum_{n=1}^{\infty}\cos(  k_{yz}^{(\alpha)}L_n^{(1)})\times K_0(L_n^{(1)}/\xi^{(\alpha)})\ ,\\
&{g} _{2}^{(1\alpha)}=\frac{4}{  k_{yz}^{(\alpha)}L_1}\sum_{n=1}^{\infty}\cos(  k_{yz}^{(\alpha)}L_n^{(2)})\times K_0(L_n^{(2)}/\xi^{(\alpha)})\ , 
\end{split}
\end{equation}
with $\alpha$ the band index and $k_{yz}^{(\alpha)}$ the in-plane Fermi momentum. $L_{\vec n}^{1,2}=2\sqrt{L_1^2n_1^2+L_2^2n_2^2},\ L_n^{(1)}=2nL_1,\ L_n^{(2)}=2nL_2,\ n,\ n_1,\ n_2\in\mathbb{N}$ are the lengths of the periodic orbits. $J_0(x)$ is the Bessel function of the first kind and $K_0(x)$ is the modified Bessel function of the first kind which suppresses the contribution of orbits longer than the superconducting coherence length in the $yz$-plane, $\xi^{(\alpha)}$.
Therefore, replacing $g_{2D}^{(\alpha)}$ by $\tilde g_{2D}^{(\alpha)}$ in the equations obtained for an infinite thin film, we simulate a finite lateral size, comparable but still larger than the thickness.

\subsection{Quantum and thermal fluctuations}\label{sec:fluctuations}
The mean-field formalism that we use is only applicable for sufficiently large systems for which quantum and thermal fluctuations are negligible. In the case of a thin film with infinite lateral size, quantum fluctuations due to size effects are negligible. At finite temperature, experimental results \cite{Zhang2010,Eom2006,Qin2009} seem to be well described by a mean-field theory even in the limit of few monolayers. This is, at first glance, surprising because, at least in the strictly two-dimensional limit, it is expected that at finite temperature there is a Kosterlitz-Thouless transition due to vortex anti-vortex unbinding. A reason for this unexpected behavior might be that the coupling to the substrate increases the effective system dimensionality. However, this must still be considered an open problem. Here we take a conservative approach and present results for thin films of at least several monolayers where it is expected, especially taking into account the coupling of the substrate, that a mean-field approach is applicable. \\
As the finite lateral size enters the nanoscale region, the thin film becomes effectively a zero-dimensional grain.  At very low temperatures ($T\ll T_c$), the deviations from mean-field predictions caused by quantum fluctuations can be neglected when the mean level spacing $\delta$ is smaller than the BCS bulk energy gap $\delta/\Delta_{bulk} \ll 1$ \cite{matveev1997,Yuzbashyan2005}. At finite temperature, thermal fluctuations smear out the superconducting phase transition in a region of temperatures $\gamma T_c$, with $\gamma =\sqrt{\delta/k_BT_c}$, around the bulk $T_c$.\cite{Muhlschlegel1972} We restrict the range of lateral sizes so that these deviations from the mean-field predictions are negligible. 

\section{Results}
In this section, we employ the theoretical formalism developed previously in order to study the interplay between shape resonances and shell effects that, in some cases, enhance superconductivity. We also investigate the influence of the coupling to the substrate and the multi-band structure that tend to suppress these size effects. \\
We present explicit results for the evolution of superconductivity in a two-band thin film as a function of the thickness, including also the coupling to the substrate. First we report results on the differences between one and two bands, the dependence of $T_c$ on the coupling constant, and the band structure parameters. We then investigate the role of shell effects that occurs when the lateral size becomes comparable to the thickness. Most results correspond to \mgb, but we also explore a broader range of parameters (see below) in order to clarify whether in realistic situations it is feasible to observe an enhancement of superconductivity due to shape resonances.\\
The coupling to the substrate is modeled by a finite step potential of height ($V_0$), which corresponds to the difference between the substrate and the thin-film chemical potential. Moreover we assign a phenomenological finite lifetime to all of the states $\tau = \gamma + \beta a$, where $a$ is the thickness and the parameters $\beta$ and $\gamma$ are estimated from recent experimental results in \mgb thin films \cite{Zhang2013}. \\
As was mentioned previously, $V_0$ and $\tau$ modify the density of states in the superconductor and therefore are important to understand its role to suppress size effects.

 The effective masses that enter in quadratic dispersion relation for each band, calculated from $m_{i\alpha}=|\partial^2 E^{(\alpha)}/\partial k_i^2|$ where $E^{(\alpha)}$ is the full energy band for MgB$_2$,\cite{Kong2001} are, in units of the electron mass,  
\begin{equation}\label{masses1}
\begin{split}
&m_{1\sigma}=3.27,\ \ m_{2\sigma}=m_{3\sigma}=0.28\ ,\\
&m_{1\pi}=0.33,\ \ m_{2\pi}=m_{3\pi}=1.00\ . 
\end{split}
\end{equation}
The constant $e_{0\pi}$ is set to different values as a way to study the influence of the band structure on superconductivity. The Debye temperature in \mgb is $\theta_D=1050$ K, which corresponds to a Debye energy $E_D = \hbar \omega_D = 90.48$ meV. 
The factors $J_{\alpha,\beta}g_{2D}^\beta/2aL^2$ in Eq. (\ref{TcT}) and $a J_{\alpha,\beta}g_{2D}^\beta K_{\alpha\beta}/8L^2$ in Eq. (\ref{TcD}) were fixed such that the solution in the bulk limit is the \mgb critical temperature $T_c \approx 38.01$ K. 
Finally, we use the following set of coupling constants \cite{Karol2006}:
\begin{equation}\label{polish_constants}
\begin{split}
&g_\sigma=0.149\ \mbox{eV}^{-1}\ g_\pi=0.29\ \mbox{eV}^{-1},\\
&\tilde V_{\sigma,\sigma}=0.694\ \mbox{eV}\ \tilde V_{\pi,\pi}=0.056\ \mbox{eV},\\
&\tilde V_{\sigma,\pi}=\tilde V_{\pi,\sigma}=0.353\ \mbox{eV},\\
&\lambda_{\sigma\sigma}=0.206\ \ \lambda_{\pi\pi}=0.033\ \ \lambda_{\sigma\pi}=0.205\ \ \lambda_{\pi\sigma}=0.105.
\raisetag{40pt}
\end{split}
\end{equation}
  In Sec. \ref{sec:coup_const}, we employ another set of $\lambda_{\alpha\beta}$ in order to study the dependence of size effects on the electron-phonon coupling.
  
\subsection{Influence of the band structure on the shape resonances of a two-band thin film}
 In this section, we analyze in detail the influence of the band structure parameters on the shape resonances observed in a two-band thin film of infinite lateral size. As it has been explained previously \cite{Thompson1963}, the superconducting properties of thin films show a non monotonous dependence with the thickness. A sawlike dependence is observed for one-band thin films where the peaks are located at values of the thickness for which a new energy subband of allowed states is occupied. Once such state is occupied, the spectral density decreases and the critical temperature drops as the thickness increases, until the following empty state can be filled. If two conduction bands are present, the same mechanism applies to each one separately. Therefore, the shape resonances pattern in the two-band case is presumably more complex or intricate than that of a one-band film depicted in Fig. \ref{Blatt_gap}.  
 
\begin{figure}[t]
\hspace{1cm}
\includegraphics[scale=0.8]{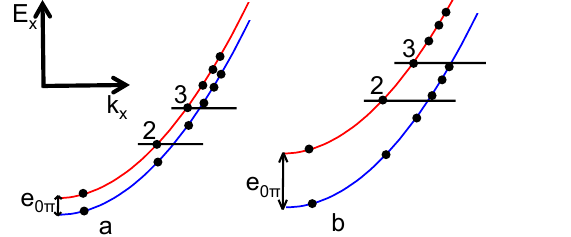}
\caption{Sketch of the dispersion relation for the $\sigma$ band (blue) and $\pi$ band (red). The number of available states in the $\sigma$ band between two consecutive states of the $\pi$ band increases as $e_{0\pi}$ increases.}
\label{parabolas}
\end{figure}

According to Eq. (\ref{bands}), the parameters that control the dispersion relation are the offset between the bands $e_{0\pi}$ and the effective masses $m_{1\alpha}$. 
As $e_{0\pi}$ is slowly increased, the number of smaller peaks, corresponding to subbands in the $\pi$ band not present in the one-band case,  is expected to increase. This behavior is straightforward to explain by simple inspection of the two dispersion relations; see Fig. \ref{parabolas}. 
We observe that, as the energy increases, states in the two bands become closer in energy. At the same time, for larger $e_{0\pi}$ [see Fig. \ref{parabolas}(b)], the number of states in the $\sigma$ band (blue) between two consecutive states of the $\pi$ band (red), labeled ``$2$'' and ``$3$'' in the figure, is larger than for smaller $e_{0\pi}$; see Fig. \ref{parabolas}(a). Therefore, as $e_{0\pi}$ increases, there are more occupied states in the $\sigma$ band before the next state in the $\pi$ band is filled.

Furthermore, as $m_{1\sigma}$ and $m_{1\pi}$ decrease, the discrete energy states are less closely packed. Therefore when a new state is occupied the change in the chemical potential is larger. This produces larger shape resonances in $T_c$. 

\begin{figure}[t]
\hspace{-0.9cm}
\includegraphics[scale=1.6]{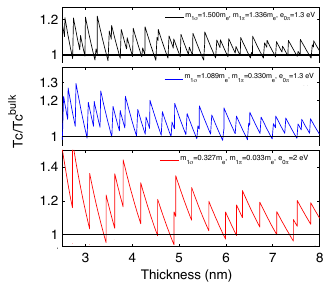}
\vspace{-0.5cm}
\caption{$T_c$ in units of $T_c^{bulk}=38.0$ K as a function of the film thickness for two-band free-standing films [Eq. (\ref{TcT})] and different effective masses. The rest of the parameters are those of \mgb [Eq.(\ref{polish_constants})]. In order to observe more clearly the shape resonances, we show the region between 2.5 and 8 nm. The in-plane effective masses are set to $m_{3\alpha}$, $m_{2\alpha}$ in Eq. (\ref{masses1}), while $e_{0\pi}$ and $m_{1\alpha}$ are indicated in each figure. Band parameters not only change the position of the shape resonances' pattern but also their amplitude.}
\label{Tc_influence}
\end{figure}

Results for the critical temperature, depicted in Fig. \ref{Tc_influence}, are fully consistent with this picture. Shown in black and blue are the oscillations in $T_c$ for different effective masses and the same $e_{0\pi}=1.3$ eV. As was expected, the shape resonances (blue) for $m_{1\sigma}=1.089m_e,\ m_{1\pi}=0.330m_e$ are slightly larger than those (black) for $m_{1\sigma}=1.500m_e,\ m_{1\pi}=1.336m_e$. Moreover, in agreement with the theoretical prediction, we observe that as $e_{0\pi}$ increases (red line) more peaks around the one corresponding to the one-band case start to be observed. 

To summarize, the band structure of the film plays an important role not only in the pattern of the shape resonances, but also in their amplitude. 

\vspace{-0.7cm}
\subsection{Differences between one and two band}
In the previous section, we have studied the intricate pattern of shape resonances observed in two-band superconducting films. In this section we compare it to the one observed in a one-band thin film with similar parameters. 

The one-band case can be recovered in two ways: the first, in which we are not interested, corresponds to the limit $e_{0\pi}\rightarrow\infty$, i.e., there is only one band available. Here we focus instead in the situation in which there are occupied states with similar energies in both bands. Provided that $e_{0\pi}=0$, we obtain states with identical quantized energies simply by setting $m_{1\sigma}=m_{1\pi}=3.27m_e$.
 \begin{figure}[t]
\includegraphics[scale=0.9]{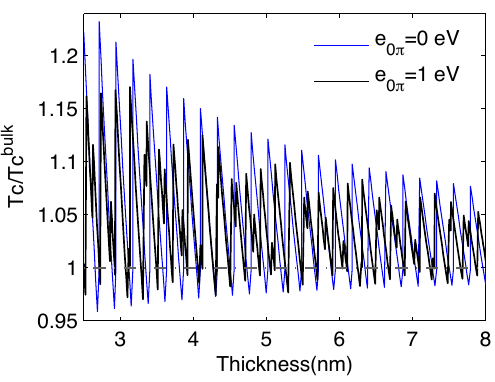}
\vspace{-0.3cm}
\caption{$T_c$ in units of $T_c^{bulk}=38.0$ K as a function of the film thickness for a one-band (blue) and a two-band (black) free-standing thin film. The parameters are those of \mgb Eq. (\ref{polish_constants}) with $m_{1\sigma}=m_{1\pi}=3.27m_e$ and in-plane masses $m_{2\alpha}$, $m_{3\alpha}$ from  Eq. (\ref{masses1}). The offset value $e_{0\pi} = 0$ corresponds to the one-band limit. The pattern of the shape resonances becomes more regular as the offset decreases and the the masses become more similar. In the one-band limit, the shape resonances have larger amplitude than in the two-band case (black). Therefore, multiband structure suppresses size effects.}
\label{Tc_diff}
\end{figure}
Using the free-standing model introduced in Sec. \ref{sec:free_model}, we obtain more regular shape resonances, depicted in Fig. \ref{Tc_diff}, than those for $m_{1\sigma}\neq m_{1\pi}$, depicted in Fig. \ref{Tc_influence}. In Fig. \ref{Tc_diff}, we compare the case of $e_{0\pi}=1$ eV (black line) with $e_{0\pi}=0$ (blue line). In the latter case the quantized components of the momentum are identical in both bands, which results in the same sawlike pattern as in the one-band superconducting film shown in Fig. \ref{Blatt_gap}. By contrast, for the reasons given in the previous section, the oscillating pattern in the two-band case has a more complex distribution of maxima and minima. Furthermore, the amplitude of the shape resonances is also smaller than in the one-band limit. This indicates that finite-size effects in two-band superconducting films are smaller than in the one-band case.

\vspace{-0.5cm}
\subsection{Role of the substrate in an infinite two-band thin film}\label{role_substrate}
Once the shape resonances in the critical temperature of a two-band superconducting free-standing film have been studied, we address the problem of the substrate influence by using the model introduced in Sec. \ref{sec:role}--\ref{sec:mat_elem}. We compute the critical temperature $T_c$ and chemical potential $\mu$ as a function of the thickness and compare them to those corresponding to a free-standing film. We restrict to infinite lateral size and thicknesses in the window [$2$,$12$] nm, a region for which recent experimental results suggest that the mean field approximation holds reasonably well.

The substrate is modeled by two parameters: the height of the step function $V_0$, namely, the mismatch between the bulk Fermi levels
of the film and substrate, and the phenomenological quasiparticle lifetime $\tau$. The first determines the eigenstate extension out of the film. The smaller the $V_0$, the larger the leaking of probability outside the film. The second parameter controls the broadening of the energy levels. We have chosen $V_0$ between $0.9$ and $1.9$ eV above the bulk film Fermi energy. This is the typical mismatch found for example in Pb films grown over a Si substrate \cite{Kirchmann2010}. \\ 
 The quasiparticle lifetime $\tau$ smoothes the shape resonances and decreases their amplitude. Since quasiparticles reach the film/substrate interface more frequently the thinner the film is, it is expected tunneling to be stronger as the thickness decreases. More specifically we expect a linear dependence with the thickness. Based on this fact and on the recent (see Fig. 3 of Ref. \cite{Zhang2013})  experimental scattering rate $\Gamma$ ($\tau=2\hbar/\Gamma$) results in \mgb thin films we propose a phenomenological expression for $\tau \approx(c_1+c_2a)$ where $a$ is the film thickness and $c_1=44.76$ fs, $c_2=0.83$ fs nm$^{-1}$ are obtained from the experimental results of  Ref. \cite{Zhang2013} between $6$ and $14$ nm. Even though the scattering rates in Ref.\cite{Zhang2013} are attributed to the film granularity, tunneling into the substrate is expected to also contribute to the level broadening. In any case, decoherence of any form is effectively modelled by a finite $\tau$ so our results are, at least qualitatively, applicable to more general situations.  
  
  As is explained in Secs. \ref{sec:role} and \ref{sec:mat_elem}, in order to obtain momentum-independent matrix elements, we approximate the interaction between all of the states by that between eigenstate whose energy is closer to being equidistant from the highest and the lowest occupied energy level.
  
\begin{figure}[t]
\includegraphics[scale=1.5]{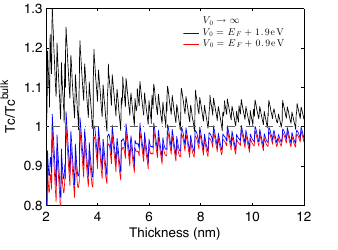}
\vspace{-0.7cm}
\caption{$T_c$ in units of $T_c^{bulk}=38.0$ K as a function of the film thickness for different couplings to the substrate, given by Eq. (\ref{TcD}) in Sec. \ref{sec:mat_elem}. $e_{0\pi}=0.05$ eV, the effective masses are given in Eq. (\ref{masses1}) and the coupling constants are given in Eq. (\ref{polish_constants}). The lifetime is $\tau\rightarrow\infty$ (black) for the free standing film while (red and blue) $\tau\mbox{(fs)}=c_1+c_2 a$, $c_1=44.76$ fs, $c_2=0.83$ fs nm$^{-1}$ and $a$ in nm. These parameters were obtained by fitting the data from $6$ to $15$ nm in Fig. $3$ of Ref. \cite{Zhang2013}. In the region of strong coupling to the substrate (blue and red lines), shape resonances are only slightly smoothed, but a significant suppression relative to the free-standing limit (black) is observed. Much smaller values of the lifetime are needed for a substantial smoothing of the shape resonances.}
\label{Tc_6}
\end{figure}

We are now ready to analyze size effects in a two-band infinite thin film for three different couplings to the substrate. In Fig. \ref{Tc_6} we depict the dependence of $T_c$ on the thickness for various values of $V_0$. It is clearly observed that shape resonances are smaller in amplitude as both the quasiparticle lifetime $\tau$ and $V_0$ decrease. Shape resonances are not substantially smoothed by the finite $\tau$ estimated from experimental results.\cite{Zhang2013} The reason for that is that the energy associated to a finite lifetime, $\Gamma \sim \hbar/\tau$, is still much smaller than the mean spacing of energy levels in the one-dimensional potential that describes confinement in the direction perpendicular to the film. Typical lifetimes of a few femtoseconds are needed to substantially smooth out the peaks and fully suppress size effects for thicknesses $\sim 5$ nm.

 In summary, as $V_0$ or $\tau$ decreases, the substrate becomes more important and any enhancement of superconductivity due to size effects is severely suppressed. We note that even the small enhancement observed in certain cases is likely not to be observable for materials for which  the surface charge neutrality condition fully applies. 

\begin{figure}[t]
\includegraphics[scale=1.51]{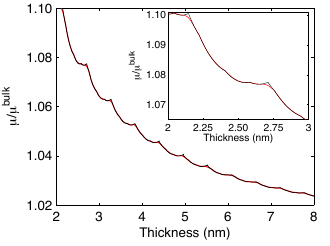}
\vspace{-0.3cm}
\caption{$\mu$ in units of $\mu^{bulk} = 3.6$ eV as a function of the film thickness. Free-standing (black) and substrate (red) of height $V_0=E_F+0.9$ eV (red). The lifetime  is $\tau\rightarrow\infty$ (black) for the free-standing, film while (red and blue) $\tau\mbox{(fs)}=c_1+c_2 a$, $c_1=44.76$ fs, $c_2=0.83$ fs nm$^{-1}$, and $a$ in nm. Masses are given by Eq. (\ref{masses1}) and $e_{0\pi}=0.05$ eV. For the free-standing film [Eq. (\ref{muT})], sharp shape resonances are clearly observed. Once the film is coupled to the substrate [Eq. (\ref{muD})], shape resonances become smoother. In the inset, smaller peaks, corresponding to the occupation of states in one band, are observed between two larger ones corresponding to the filling of states in the other band.}
\label{mu_6}
\end{figure}

Shown in Fig. \ref{mu_6} are the shape resonances in the chemical potential for a thickness in the same region as in Fig. \ref{Tc_6}. The overall magnitude of $\mu$ shows no significant difference compared to the free-standing limit. However (see inset of Fig. \ref{mu_6}), the pattern is slightly smoothed. As in the case of $T_c$, a smaller $\tau$ results in a smoother behavior which becomes monotonically decreasing for sufficiently small lifetime.
\vspace{-0.5cm}
\subsection{Influence of the electron-phonon coupling constants}\label{sec:coup_const}
In this section, we study size effects for different electron-phonon coupling constants and fixed band structure parameters. We take the effective masses given in Eq. (\ref{masses1}), $e_{0\pi}=0.05$ eV, $V_0=E_F+0.9$ eV, and $\tau\mbox{(fs)}=c_1+c_2 a$, $c_1=44.76$ fs, $c_2=0.83$ fs nm$^{-1}$, and $a$ in nm.

\begin{figure}[t]
\center
\includegraphics[scale=1.4]{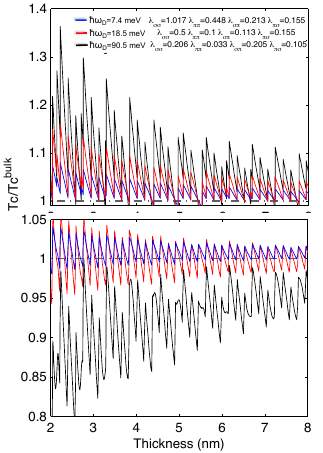}
\caption{$T_c$ in units of $T_c^{bulk}\approx38.0$ K as a function of the film thickness for different values of the electron-phonon coupling constant. Upper: free standing film, Eq. (\ref{TcT}). Lower: substrate, from Eq. (\ref{TcD}), included. Masses in all cases are given by Eq. (\ref{masses1}) and $e_{0\pi}=0.05$ eV. The Debye energy is tuned so that in all cases, $T_c^{bulk} \approx 38.0$ K. As the coupling constant decreases, finite-size effects are clearly stronger, though the suppression due to the substrate is also stronger. The optimal setting results from a delicate balance between these two factors. }
\label{Tc_other_coup_const}
\end{figure}

Black lines in Fig. \ref{Tc_other_coup_const} correspond to the coupling constant employed in previous sections [Eq. (\ref{polish_constants})] and are also shown in Fig. \ref{Tc_6}. Shown in blue are the results corresponding to the coupling constants from Ref. \cite{Golubov2002} and a Debye energy $\hbar\omega_D=7.4$ meV \cite{Liu2001} that gives $T_c^{bulk}=38.3$ K. Red lines correspond to a set of coupling constants and a Debye energy, not related to MgB$_2$, but with the same bulk critical temperature $T_c^{bulk}=38.2$ K. It is clearly observed (see Fig. \ref{Tc_other_coup_const}) that larger coupling constants lead to weaker finite-size effects and less suppression of $T_c$ with respect to the bulk limit. This follows straightforwardly from Eqs. (\ref{anu}) and (\ref{gapTh}) by calculating the first order correction to $\Delta$ which is inversely proportional to the dimensionless coupling constant. Therefore, a larger coupling constant leads to smaller finite size effects.

\vspace{-0.5cm}
\subsection{Finite lateral size and shell effects}
In this section we study the role of a finite lateral size in the two-band thin films studied previously. In order to neglect thermal fluctuations, which are beyond the mean-field approximation, we restrict to lateral sizes of the order of, but larger than the film thickness $\sim 10$nm. Technically, the first consequence of a finite lateral size is that the integrals over $k_y$ and $k_z$ in the gap equations have to replaced by discrete sums. Moreover, due to the isotropic in-plane effective masses and assuming a square shape, we expect level degeneracy, namely, several states occupy the same energy level, usually referred to as a shell. This bunching of levels induces larger fluctuations in the spectral density, i.e., the so-called shell effects, that are also expected to have an important impact on the superconducting properties of the material.\cite{parmenter,aggalt2008,shanenko2007prb}.

In Sec. \ref{sec:finite_lateral_size}, we discussed that shell effects can be analytically included by simply replacing the bulk two-dimensional density of states in each band, $g^{(\alpha)}_{2D}=m^{(\alpha)}_{2D}L^2/\pi\hbar^2$ by $\tilde g^{(\alpha)}_{2D}\simeq g^{(\alpha)}_{2D}[1+\overline{g}^{(\alpha)}+g ^{(\alpha)}_l]$. The latter only depends on the lateral size, in-plane coherence lengths $\xi^{(\alpha)}=\xi^{(\alpha)}_{yz}$, and the in-plane Fermi momentum $k^{(\alpha)}_{yz}$. Therefore, $\tilde g_{2D}$ is constant in energy for all films with the same lateral size. As a consequence we can replace $g_{2D}^{(\alpha)}$ by $\tilde g_{2D}^{(\alpha)}$ in Eq. (\ref{TcD}).
 
In order to get explicit results  we use \mgb\hspace{-0.15cm} parameters. For the in-plane coherence lengths, we take $\xi^{(\sigma)}_{yz}=13$ nm, $\xi^{(\pi)}_{yz}=51$nm (at $T=0$),\cite{Moshchalkov2009} while a simple calculation of the in-plane Fermi momenta yields $k_{yz}^{(\alpha)}=\sqrt{m_{2\alpha} m_{3\alpha}} v_{yz}^{(\alpha)}/\hbar$. These are the components corresponding to the crystallographic $ab$-plane of the \mgb cell, which is the $yz$-plane in our coordinate system. The effective masses are given in Eq. (\ref{masses1}) and the in-plane components of the Fermi velocities $v_{yz}^{(\alpha)}$ are taken from Ref.\cite{Brinkman2002}, $v_{yz}^{(\sigma)}=4.40\times10^5$ m/s and $v_{yz}^{(\pi)}=5.35\times10^5$ m/s,
\begin{equation}\label{Fermi_level}
 k_{yz}^{(\sigma)}=1.0710\ \text{nm}^{-1},\ \  k_{yz}^{(\pi)}=4.6311\ \text{nm}^{-1}.
\end{equation}
As  was mentioned in Sec. \ref{sec:fluctuations}, a mean-field approach is only valid for sizes in which fluctuations are not important which, for thermal fluctuations, depends on the ratio between the mean level spacing and $T_c$. For conventional superconductors, a lateral size and thickness of at least $\sim 10$ and $\sim 5$ nm, respectively, are typical requirements for a mean-field formalism to be applicable. 
For \mgb first-principles calculations \cite{Kong2001} suggest that the density of states at $E_F$ in each band is $N_{\sigma}(E_F)=0.150$ states/(eV-spin-cell) and  $N_{\pi}(E_F)=0.205$ states/(eV spin cell). Using the unit cell parameters \cite{Nagamatsu2001}  $a=3.086$ \AA\ (not to be confused with the film thickness) and $c=3.524$ \AA\  the mean level spacing for each band is,  $\delta_\sigma=\frac{0.097}{\mathcal{V}} \text{ eV, }\ \ \  \delta_\pi=\frac{0.070}{\mathcal{V}} \text{ eV}$, where $\mathcal{V}$ is the film volume in $nm^3$. 
 For an isolated film of thickness $a=6$ nm and lateral size $L=12\times12$ nm$^2$ at $T_c=38$ K the magnitude of thermal fluctuations is controlled by the parameter,
\begin{equation*}
\sqrt{\frac{\delta_\sigma}{k_BT_c}}\simeq 0.19,\ \ \ \sqrt{\frac{\delta_\pi}{k_BT_c}}\simeq0.16.
\end{equation*}
 For free-standing films, this is the typical minimum size for which thermal fluctuations are negligible and a mean-field approach is applicable. We expect that the presence of the substrate reduces fluctuations induced by size effects. However, we take a conservative stance and restrict our study to volumes $\geq6\times12\times12$ nm$^3$.
\begin{figure}[H]
\hspace{-0.5cm}
\includegraphics[height=7cm,width=9cm]{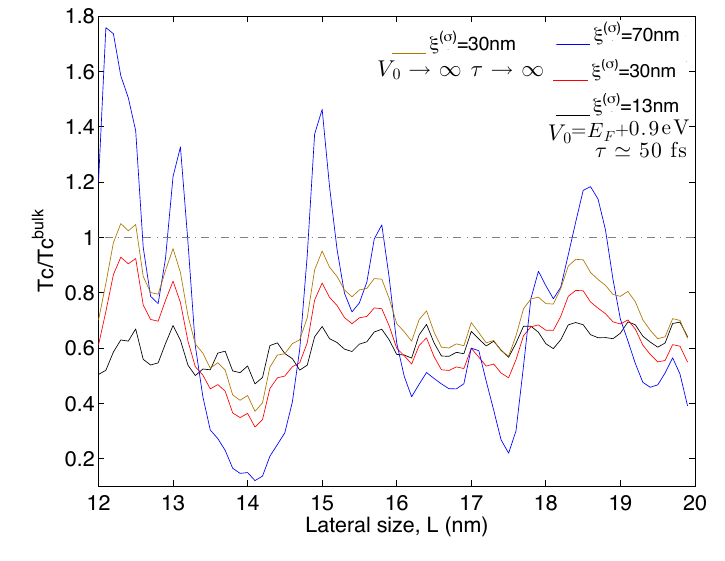}
\vspace{-0.2cm}
\caption{$T_c$  in units of $T_c^{bulk}=38.0$ K as a function of the lateral size for a thickness equal to $6.16$ nm. As in the previous figures the band parameters are those of MgB$_2$. We employ Eq. (\ref{TcD}) but replace  $g_{2D}^{(\alpha)}$ by $\tilde g_{2D}^{(\alpha)}$, given by Eqs. (\ref{factors1})--(\ref{factors3}), to study shell effects for different in-plane coherence lengths in the $\sigma$ band. In the other band, $\xi_{\pi}=51$ nm and masses can be found in Eq. (\ref{masses1}). For the range of thicknesses in which thermal fluctuations are not important, and including the coupling to the substrate, shell effects enhance superconductivity only for coherence lengths considerably larger (by a factor of two) than the film lateral size. In the free-standing film limit (yellow line) a moderate enhancement is observed for $L\sim12$ nm. For a substantial enhancement of superconductivity, the coherence length of the material must be much larger than that of MgB$_2$.}
\label{Tc_finite}
\end{figure}

Results depicted in Fig. \ref{Tc_finite} show that for a finite lateral size $\sim12$nm, shell effects induce corrections in $T_c$ which are much stronger than those of a thin film with a finite thickness of the same order and infinite lateral size as seen in Fig. \ref{Tc_6}. For \mgb (in black), no enhancement of superconductivity with respect to the bulk limit is observed. This is due to the small coherence length in the $\sigma$ band of MgB$_2$, compared to the lateral size. In this situation, the oscillating terms $g _l^{(\alpha)}$ are suppressed by the modified Bessel functions $K_0$, given by Eq. (\ref{factors3}). As a result the leading correction is $\overline g_{2D}^{(\alpha)}$, given by Eq. (\ref{factors2}), which is negative. Therefore, $g_{2D}^\sigma\approx g_{2D}^\sigma[1+\overline g_{2D}^{(\alpha)}]<g_{2D}^\sigma$, given by Eq. (\ref{factors1}) and superconductivity is suppressed  by a finite lateral size. In the limit $L\to\infty$, we recover the  infinite lateral size result $T_c(L\to\infty)$.

In the case of a superconducting coherence length (blue line) much larger than the lateral size, we observe a substantial enhancement of the critical temperature. This is a consequence of shell effects in the two dimensional spectral density that are not smoothed out by a small coherence length.   
\subsection{Limitations and limits of applicability of the model}
We briefly review the limits of applicability of the results and the different approximations that we employ across the paper.

The mean-field approach that we employ neglects quantum and thermal fluctuations. As was mentioned previously, this is a good approximation for sufficiently large lateral sizes, though we note that even for an infinite lateral size we expect that the mean-field approach breaks down in the strictly two dimensional limit where a Kosterlitz-Thouless transition occurs at a lower temperature. However results from recent experiments \cite{xue2004science,xue2010natphys,shih2009science} in Pb ultra-thin films that explore the two dimensional limit were, at least qualitatively, well described by a mean-field formalism. A reason for that behavior is that the substrate increases the effective dimensionality of the system and consequently suppresses the Kosterlitz-Thouless transition. Since this issue is not yet settled here we have opted to present results only for thicknesses of at least  $2$nm where a mean field formalism should still be applicable.   

The coupling to the substrate is modeled by a phenomenological quasiparticle lifetime to describe tunneling into the substrate and a step potential to describe the substrate thin-film interface. A more realistic model of the tunneling mechanism, beyond the scope of this paper, requires a much more detailed knowledge of the interface which depends on the growth techniques and the material substrate. 

We have used the zero temperature coherence length of MgB$_2$. However estimations\cite{Ozer2006} of the coherence length in Pb film show substantial changes in the coherence length for different system sizes. This coherence length is an input in our model so that once the coherence length in nanoscale samples is known the calculation of $T_c$ could easily be updated accordingly.

We do not consider the full band dispersion relation but rather we have expanded it up to second order around the Fermi level. This approximation might neglect some non-trivial influence of the bands specially in observables, such as the conductivity, which involve energies substantially larger than the gap. However, we expect this approximation to be fair in the calculation of quantities such as $T_c$ and the superconducting gap that involves energies close to the Fermi energy. 

We have considered crystalline films in the absence of impurities or strain due to lattice mismatch with the substrate. Current state of the art experimental techniques are capable of manufacturing samples with these properties. 

\section{Conclusions}
We have investigated analytically the evolution of superconductivity, including the coupling to the substrate, in multi-band thin-films as the thickness and lateral size enter the nanoscale region. \\Shape-resonances in two-band thin films, neglecting the substrate, are more irregular and lead to a more modest enhancement of superconductivity than in one-band films. Size effects are stronger as the effective electron-phonon coupling is decreased. Qualitatively similar results are obtained for different effective masses describing the band structure though smaller masses tend to induce stronger size effects. We have observed that a finite lateral size $\sim 10$nm induces additional size effects, i.e., the so-called shell effects, which can enhance superconductivity in materials in which the coherence length is much longer than the lateral size. For smaller lateral sizes, thermal fluctuations, not included in our model,  become important and our results are not reliable. 
\\
Once the substrate is considered the average enhancement is strongly suppressed. As thickness is decreased, tunneling is expected to be more important, smoothing the pattern of shape resonances. However, in the range of parameters used, this smoothing is rather weak. The critical temperature and the amplitude of shape resonances decrease as well.  The case of MgB$_2$, a two-band superconductor, is discussed in detail. In the relatively broad range of parameters that we  explore we did not observe a substantial enhancement of superconductivity once the multi-band structure and the substrate are considered simultaneously. It is likely that even this modest enhancement of $T_c$ is not observable for materials in which the charge neutrality condition applies. 
\section{Acknowledgments}
AR acknowledge financial support from a ``laCaixa" Foundation fellowship. AMG was supported by EPSRC, Grant No. EP/I004637/1, FCT, Grant No. PTDC/FIS/111348/2009, and
a Marie Curie International Reintegration Grant No.
PIRG07-GA-2010-268172.
\appendix
\section{}

\numberwithin{equation}{section}
\subsection{Poisson summation formula}
Given $f$ as non-negative, decreasing, and continuous on $[0,\infty)$ and that $\lim_{b\to\infty}\int_0^b f(x) dx$ exists, then
\begin{equation}\label{Poiss}
\hspace{-1mm}\sqrt{\sigma}\left[\frac{1}{2}f(0)+\sum_{n=1}^\infty f(n\sigma)\right)=\sqrt{\lambda}\left(\frac{1}{2}h(0)+\sum_{l=1}^\infty h(l\lambda)\right],
\end{equation}
where $\sigma\lambda=2\pi$ and $h(y)=\sqrt{2/\pi}\int_0^\infty f(m)\cos(my)dm$ \cite{Apostol}. Setting $\sigma=1$, $\lambda=2\pi$ and defining $f(n)=\mu-\eta^\alpha(n)$, where  $\eta^\alpha(n)=e_{0\alpha}+E_0^\alpha n^2$ we substitute Eq. (\ref{Poiss}) into Eq. (\ref{NT}). To simplify notation, we omit the band index  $\alpha$. 
\onecolumngrid
$f(n)$ satisfies the necessary conditions to use Eq. (\ref{Poiss}) when $\eta\in[e_0,\mu]$. Thus, integrating in energy between $e_0$ and $\mu$ and restricting the sum on the left-hand side from $n=1$ to $\nu$, the Poisson summation formula leads to
\begin{equation}\label{mu_Poiss}
\frac{\mu-e_0}{2}+\sum_{n=1}^\nu\left(\mu-\eta_n\right)=\frac{2}{3\sqrt{E_0}}(\mu-e_0)^{3/2}+\sum_{l=1}^\infty\left[-\frac{\sqrt{E_0(\mu-e_0)}}{\pi^2 l^2}\cos\left(2\pi l \sqrt{\frac{\mu-e_0}{E_0}}\right)+\frac{E_0}{2\pi^3 l^3}\sin\left(2\pi l \sqrt{\frac{\mu-e_0}{E_0}}\right)\right]\ .
\end{equation}
\subsection{Factors $K_{\alpha\beta}$}
Here we present factors from the interaction matrix elements. $k_{m_\alpha}$ and $\kappa_{m_\alpha}$ are defined in Sec. \ref{sec:mat_elem}.
\begin{equation}\label{K_alpha_beta}
K_{\alpha\beta}=\frac{1}{\frac{a}{2}-\frac{\sin(2k_{m_\alpha}a)}{4\kappa_{m_\alpha}}+\frac{\sin^2(2k_{m_\alpha}a)}{2\kappa_{m_\alpha}}}\frac{1}{\frac{a}{2}-\frac{\sin(2k_{m_\beta}a)}{4\kappa_{m_\beta}}+\frac{\sin^2(2k_{m_\beta}a)}{2\kappa_{m_\beta}}}\ .
\end{equation}
\twocolumngrid
\bibliography{biblio}

\begin{thebibliography}{42}%
\makeatletter
\providecommand \@ifxundefined [1]{%
 \@ifx{#1\undefined}
}%
\providecommand \@ifnum [1]{%
 \ifnum #1\expandafter \@firstoftwo
 \else \expandafter \@secondoftwo
 \fi
}%
\providecommand \@ifx [1]{%
 \ifx #1\expandafter \@firstoftwo
 \else \expandafter \@secondoftwo
 \fi
}%
\providecommand \natexlab [1]{#1}%
\providecommand \enquote  [1]{``#1''}%
\providecommand \bibnamefont  [1]{#1}%
\providecommand \bibfnamefont [1]{#1}%
\providecommand \citenamefont [1]{#1}%
\providecommand \href@noop [0]{\@secondoftwo}%
\providecommand \href [0]{\begingroup \@sanitize@url \@href}%
\providecommand \@href[1]{\@@startlink{#1}\@@href}%
\providecommand \@@href[1]{\endgroup#1\@@endlink}%
\providecommand \@sanitize@url [0]{\catcode `\\12\catcode `\$12\catcode
  `\&12\catcode `\#12\catcode `\^12\catcode `\_12\catcode `\%12\relax}%
\providecommand \@@startlink[1]{}%
\providecommand \@@endlink[0]{}%
\providecommand \url  [0]{\begingroup\@sanitize@url \@url }%
\providecommand \@url [1]{\endgroup\@href {#1}{\urlprefix }}%
\providecommand \urlprefix  [0]{URL }%
\providecommand \Eprint [0]{\href }%
\providecommand \doibase [0]{http://dx.doi.org/}%
\providecommand \selectlanguage [0]{\@gobble}%
\providecommand \bibinfo  [0]{\@secondoftwo}%
\providecommand \bibfield  [0]{\@secondoftwo}%
\providecommand \translation [1]{[#1]}%
\providecommand \BibitemOpen [0]{}%
\providecommand \bibitemStop [0]{}%
\providecommand \bibitemNoStop [0]{.\EOS\space}%
\providecommand \EOS [0]{\spacefactor3000\relax}%
\providecommand \BibitemShut  [1]{\csname bibitem#1\endcsname}%
\let\auto@bib@innerbib\@empty
\bibitem [{\citenamefont {Guo}\ \emph {et~al.}(2004)\citenamefont {Guo},
  \citenamefont {Zhang}, \citenamefont {Bao}, \citenamefont {Han},
  \citenamefont {Tang}, \citenamefont {Zhang}, \citenamefont {Zhu},
  \citenamefont {Wang}, \citenamefont {Niu}, \citenamefont {Qiu} \emph
  {et~al.}}]{xue2004science}%
  \BibitemOpen
  \bibfield  {author} {\bibinfo {author} {\bibfnamefont {Y.}~\bibnamefont
  {Guo}}, \bibinfo {author} {\bibfnamefont {Y.-F.}\ \bibnamefont {Zhang}},
  \bibinfo {author} {\bibfnamefont {X.-Y.}\ \bibnamefont {Bao}}, \bibinfo
  {author} {\bibfnamefont {T.-Z.}\ \bibnamefont {Han}}, \bibinfo {author}
  {\bibfnamefont {Z.}~\bibnamefont {Tang}}, \bibinfo {author} {\bibfnamefont
  {L.-X.}\ \bibnamefont {Zhang}}, \bibinfo {author} {\bibfnamefont {W.-G.}\
  \bibnamefont {Zhu}}, \bibinfo {author} {\bibfnamefont {E.}~\bibnamefont
  {Wang}}, \bibinfo {author} {\bibfnamefont {Q.}~\bibnamefont {Niu}}, \bibinfo
  {author} {\bibfnamefont {Z.}~\bibnamefont {Qiu}},  \emph {et~al.},\ }\href
  {\doibase 10.1126/science.1105130} {\bibfield  {journal} {\bibinfo  {journal}
  {Science}\ }\textbf {\bibinfo {volume} {306}},\ \bibinfo {pages} {1915}
  (\bibinfo {year} {2004})}\BibitemShut {NoStop}%
\bibitem [{\citenamefont {Zhang}\ \emph
  {et~al.}(2010{\natexlab{a}})\citenamefont {Zhang}, \citenamefont {Cheng},
  \citenamefont {Li}, \citenamefont {Sun}, \citenamefont {Wang}, \citenamefont
  {Zhu}, \citenamefont {He}, \citenamefont {Wang}, \citenamefont {Ma},
  \citenamefont {Chen} \emph {et~al.}}]{xue2010natphys}%
  \BibitemOpen
  \bibfield  {author} {\bibinfo {author} {\bibfnamefont {T.}~\bibnamefont
  {Zhang}}, \bibinfo {author} {\bibfnamefont {P.}~\bibnamefont {Cheng}},
  \bibinfo {author} {\bibfnamefont {W.-J.}\ \bibnamefont {Li}}, \bibinfo
  {author} {\bibfnamefont {Y.-J.}\ \bibnamefont {Sun}}, \bibinfo {author}
  {\bibfnamefont {G.}~\bibnamefont {Wang}}, \bibinfo {author} {\bibfnamefont
  {X.-G.}\ \bibnamefont {Zhu}}, \bibinfo {author} {\bibfnamefont
  {K.}~\bibnamefont {He}}, \bibinfo {author} {\bibfnamefont {L.}~\bibnamefont
  {Wang}}, \bibinfo {author} {\bibfnamefont {X.}~\bibnamefont {Ma}}, \bibinfo
  {author} {\bibfnamefont {X.}~\bibnamefont {Chen}},  \emph {et~al.},\ }\href
  {\doibase 10.1038/nphys1499} {\bibfield  {journal} {\bibinfo  {journal} {Nat.
  Phys.}\ }\textbf {\bibinfo {volume} {6}},\ \bibinfo {pages} {104} (\bibinfo
  {year} {2010}{\natexlab{a}})}\BibitemShut {NoStop}%
\bibitem [{\citenamefont {Qin}\ \emph {et~al.}(2009)\citenamefont {Qin},
  \citenamefont {Kim}, \citenamefont {Niu},\ and\ \citenamefont
  {Shih}}]{shih2009science}%
  \BibitemOpen
  \bibfield  {author} {\bibinfo {author} {\bibfnamefont {S.}~\bibnamefont
  {Qin}}, \bibinfo {author} {\bibfnamefont {J.}~\bibnamefont {Kim}}, \bibinfo
  {author} {\bibfnamefont {Q.}~\bibnamefont {Niu}}, \ and\ \bibinfo {author}
  {\bibfnamefont {C.-K.}\ \bibnamefont {Shih}},\ }\href {\doibase
  10.1126/science.1170775} {\bibfield  {journal} {\bibinfo  {journal}
  {Science}\ }\textbf {\bibinfo {volume} {324}},\ \bibinfo {pages} {1314}
  (\bibinfo {year} {2009})}\BibitemShut {NoStop}%
\bibitem [{\citenamefont {Bose}\ \emph {et~al.}(2010)\citenamefont {Bose},
  \citenamefont {Garc{\'\i}a-Garc{\'\i}a}, \citenamefont {Ugeda}, \citenamefont
  {Urbina}, \citenamefont {Michaelis}, \citenamefont {Brihuega},\ and\
  \citenamefont {Kern}}]{Bose2010}%
  \BibitemOpen
  \bibfield  {author} {\bibinfo {author} {\bibfnamefont {S.}~\bibnamefont
  {Bose}}, \bibinfo {author} {\bibfnamefont {A.}~\bibnamefont
  {Garc{\'\i}a-Garc{\'\i}a}}, \bibinfo {author} {\bibfnamefont
  {M.}~\bibnamefont {Ugeda}}, \bibinfo {author} {\bibfnamefont
  {J.}~\bibnamefont {Urbina}}, \bibinfo {author} {\bibfnamefont
  {C.}~\bibnamefont {Michaelis}}, \bibinfo {author} {\bibfnamefont
  {I.}~\bibnamefont {Brihuega}}, \ and\ \bibinfo {author} {\bibfnamefont
  {K.}~\bibnamefont {Kern}},\ }\href {\doibase 10.1038/nmat2768} {\bibfield
  {journal} {\bibinfo  {journal} {Nat. Mat.}\ }\textbf {\bibinfo {volume}
  {9}},\ \bibinfo {pages} {550} (\bibinfo {year} {2010})}\BibitemShut {NoStop}%
\bibitem [{\citenamefont {Brun}\ \emph {et~al.}(2009)\citenamefont {Brun},
  \citenamefont {Hong}, \citenamefont {Patthey}, \citenamefont {Sklyadneva},
  \citenamefont {Heid}, \citenamefont {Echenique}, \citenamefont {Bohnen},
  \citenamefont {Chulkov},\ and\ \citenamefont {Schneider}}]{wolf2009}%
  \BibitemOpen
  \bibfield  {author} {\bibinfo {author} {\bibfnamefont {C.}~\bibnamefont
  {Brun}}, \bibinfo {author} {\bibfnamefont {I.-P.}\ \bibnamefont {Hong}},
  \bibinfo {author} {\bibfnamefont {F.}~\bibnamefont {Patthey}}, \bibinfo
  {author} {\bibfnamefont {I.~Y.}\ \bibnamefont {Sklyadneva}}, \bibinfo
  {author} {\bibfnamefont {R.}~\bibnamefont {Heid}}, \bibinfo {author}
  {\bibfnamefont {P.}~\bibnamefont {Echenique}}, \bibinfo {author}
  {\bibfnamefont {K.}~\bibnamefont {Bohnen}}, \bibinfo {author} {\bibfnamefont
  {E.}~\bibnamefont {Chulkov}}, \ and\ \bibinfo {author} {\bibfnamefont
  {W.-D.}\ \bibnamefont {Schneider}},\ }\href {\doibase
  10.1103/PhysRevLett.102.207002} {\bibfield  {journal} {\bibinfo  {journal}
  {Phys. Rev. Lett.}\ }\textbf {\bibinfo {volume} {102}},\ \bibinfo {pages}
  {207002} (\bibinfo {year} {2009})}\BibitemShut {NoStop}%
\bibitem [{\citenamefont {Brun}\ \emph {et~al.}(2012)\citenamefont {Brun},
  \citenamefont {M{\"u}ller}, \citenamefont {Hong}, \citenamefont {Patthey},
  \citenamefont {Flindt},\ and\ \citenamefont {Schneider}}]{wolf2012}%
  \BibitemOpen
  \bibfield  {author} {\bibinfo {author} {\bibfnamefont {C.}~\bibnamefont
  {Brun}}, \bibinfo {author} {\bibfnamefont {K.~H.}\ \bibnamefont
  {M{\"u}ller}}, \bibinfo {author} {\bibfnamefont {I.-P.}\ \bibnamefont
  {Hong}}, \bibinfo {author} {\bibfnamefont {F.}~\bibnamefont {Patthey}},
  \bibinfo {author} {\bibfnamefont {C.}~\bibnamefont {Flindt}}, \ and\ \bibinfo
  {author} {\bibfnamefont {W.-D.}\ \bibnamefont {Schneider}},\ }\href {\doibase
  10.1103/PhysRevLett.108.126802} {\bibfield  {journal} {\bibinfo  {journal}
  {Phys. Rev. Lett.}\ }\textbf {\bibinfo {volume} {108}},\ \bibinfo {pages}
  {126802} (\bibinfo {year} {2012})},\ \Eprint {http://arxiv.org/abs/1006.0333}
  {arXiv:1006.0333} \BibitemShut {NoStop}%
\bibitem [{\citenamefont {Brihuega}\ \emph {et~al.}(2011)\citenamefont
  {Brihuega}, \citenamefont {Garc{\'\i}a-Garc{\'\i}a}, \citenamefont {Ribeiro},
  \citenamefont {Ugeda}, \citenamefont {Michaelis}, \citenamefont {Bose},\ and\
  \citenamefont {Kern}}]{sangita2011prb}%
  \BibitemOpen
  \bibfield  {author} {\bibinfo {author} {\bibfnamefont {I.}~\bibnamefont
  {Brihuega}}, \bibinfo {author} {\bibfnamefont {A.~M.}\ \bibnamefont
  {Garc{\'\i}a-Garc{\'\i}a}}, \bibinfo {author} {\bibfnamefont
  {P.}~\bibnamefont {Ribeiro}}, \bibinfo {author} {\bibfnamefont {M.~M.}\
  \bibnamefont {Ugeda}}, \bibinfo {author} {\bibfnamefont {C.~H.}\ \bibnamefont
  {Michaelis}}, \bibinfo {author} {\bibfnamefont {S.}~\bibnamefont {Bose}}, \
  and\ \bibinfo {author} {\bibfnamefont {K.}~\bibnamefont {Kern}},\ }\href
  {\doibase 10.1103/PhysRevB.84.104525} {\bibfield  {journal} {\bibinfo
  {journal} {Phys. Rev. B}\ }\textbf {\bibinfo {volume} {84}},\ \bibinfo
  {pages} {104525} (\bibinfo {year} {2011})},\ \Eprint
  {http://arxiv.org/abs/0904.0354} {arXiv:0904.0354} \BibitemShut {NoStop}%
\bibitem [{\citenamefont {Thompson}\ and\ \citenamefont
  {Blatt}(1963)}]{Thompson1963}%
  \BibitemOpen
  \bibfield  {author} {\bibinfo {author} {\bibfnamefont {C.~J.}\ \bibnamefont
  {Thompson}}\ and\ \bibinfo {author} {\bibfnamefont {J.~M.}\ \bibnamefont
  {Blatt}},\ }\href {\doibase 10.1016/S0375-9601(63)80003-1} {\bibfield
  {journal} {\bibinfo  {journal} {Phys. Lett.}\ }\textbf {\bibinfo {volume}
  {5}},\ \bibinfo {pages} {6} (\bibinfo {year} {1963})}\BibitemShut {NoStop}%
\bibitem [{\citenamefont {Allen}(1975)}]{allen1975}%
  \BibitemOpen
  \bibfield  {author} {\bibinfo {author} {\bibfnamefont {R.~E.}\ \bibnamefont
  {Allen}},\ }\href {\doibase 10.1103/PhysRevB.12.3650} {\bibfield  {journal}
  {\bibinfo  {journal} {Phys. Rev. B}\ }\textbf {\bibinfo {volume} {12}},\
  \bibinfo {pages} {3650} (\bibinfo {year} {1975})}\BibitemShut {NoStop}%
\bibitem [{\citenamefont {Yu}\ \emph {et~al.}(1976)\citenamefont {Yu},
  \citenamefont {Strongin},\ and\ \citenamefont {Paskin}}]{paskin1976}%
  \BibitemOpen
  \bibfield  {author} {\bibinfo {author} {\bibfnamefont {M.}~\bibnamefont
  {Yu}}, \bibinfo {author} {\bibfnamefont {M.}~\bibnamefont {Strongin}}, \ and\
  \bibinfo {author} {\bibfnamefont {A.}~\bibnamefont {Paskin}},\ }\href
  {\doibase 10.1103/PhysRevB.14.996} {\bibfield  {journal} {\bibinfo  {journal}
  {Phys. Rev. B}\ }\textbf {\bibinfo {volume} {14}},\ \bibinfo {pages} {996}
  (\bibinfo {year} {1976})}\BibitemShut {NoStop}%
\bibitem [{\citenamefont {Shanenko}\ \emph {et~al.}(2007)\citenamefont
  {Shanenko}, \citenamefont {Croitoru},\ and\ \citenamefont
  {Peeters}}]{shanenko2007prb}%
  \BibitemOpen
  \bibfield  {author} {\bibinfo {author} {\bibfnamefont {A.}~\bibnamefont
  {Shanenko}}, \bibinfo {author} {\bibfnamefont {M.}~\bibnamefont {Croitoru}},
  \ and\ \bibinfo {author} {\bibfnamefont {F.}~\bibnamefont {Peeters}},\ }\href
  {\doibase 10.1103/PhysRevB.75.014519} {\bibfield  {journal} {\bibinfo
  {journal} {Phys. Rev. B}\ }\textbf {\bibinfo {volume} {75}},\ \bibinfo
  {pages} {014519} (\bibinfo {year} {2007})}\BibitemShut {NoStop}%
\bibitem [{\citenamefont {Chen}\ \emph {et~al.}(2006)\citenamefont {Chen},
  \citenamefont {Zhu},\ and\ \citenamefont {Xie}}]{chen2006quantum}%
  \BibitemOpen
  \bibfield  {author} {\bibinfo {author} {\bibfnamefont {B.}~\bibnamefont
  {Chen}}, \bibinfo {author} {\bibfnamefont {Z.}~\bibnamefont {Zhu}}, \ and\
  \bibinfo {author} {\bibfnamefont {X.}~\bibnamefont {Xie}},\ }\href {\doibase
  10.1103/PhysRevB.74.132504} {\bibfield  {journal} {\bibinfo  {journal} {Phys.
  Rev. B}\ }\textbf {\bibinfo {volume} {74}},\ \bibinfo {pages} {132504}
  (\bibinfo {year} {2006})},\ \Eprint {http://arxiv.org/abs/0512290}
  {cond-mat:0512290} \BibitemShut {NoStop}%
\bibitem [{\citenamefont {Gozar}\ \emph {et~al.}(2008)\citenamefont {Gozar},
  \citenamefont {Logvenov}, \citenamefont {Kourkoutis}, \citenamefont
  {Bollinger}, \citenamefont {Giannuzzi}, \citenamefont {Muller},\ and\
  \citenamefont {Bozovic}}]{bozovic2008nature}%
  \BibitemOpen
  \bibfield  {author} {\bibinfo {author} {\bibfnamefont {A.}~\bibnamefont
  {Gozar}}, \bibinfo {author} {\bibfnamefont {G.}~\bibnamefont {Logvenov}},
  \bibinfo {author} {\bibfnamefont {L.~F.}\ \bibnamefont {Kourkoutis}},
  \bibinfo {author} {\bibfnamefont {A.}~\bibnamefont {Bollinger}}, \bibinfo
  {author} {\bibfnamefont {L.}~\bibnamefont {Giannuzzi}}, \bibinfo {author}
  {\bibfnamefont {D.}~\bibnamefont {Muller}}, \ and\ \bibinfo {author}
  {\bibfnamefont {I.}~\bibnamefont {Bozovic}},\ }\href {\doibase
  10.1038/nature07293} {\bibfield  {journal} {\bibinfo  {journal} {Nature}\
  }\textbf {\bibinfo {volume} {455}},\ \bibinfo {pages} {782} (\bibinfo {year}
  {2008})},\ \Eprint {http://arxiv.org/abs/0810.1890} {arXiv:0810.1890}
  \BibitemShut {NoStop}%
\bibitem [{\citenamefont {Liu}\ \emph {et~al.}(2012)\citenamefont {Liu},
  \citenamefont {Zhang}, \citenamefont {Mou}, \citenamefont {He}, \citenamefont
  {Ou}, \citenamefont {Wang}, \citenamefont {Li}, \citenamefont {Wang},
  \citenamefont {Zhao}, \citenamefont {He} \emph {et~al.}}]{xue2012natmat}%
  \BibitemOpen
  \bibfield  {author} {\bibinfo {author} {\bibfnamefont {D.}~\bibnamefont
  {Liu}}, \bibinfo {author} {\bibfnamefont {W.}~\bibnamefont {Zhang}}, \bibinfo
  {author} {\bibfnamefont {D.}~\bibnamefont {Mou}}, \bibinfo {author}
  {\bibfnamefont {J.}~\bibnamefont {He}}, \bibinfo {author} {\bibfnamefont
  {Y.-B.}\ \bibnamefont {Ou}}, \bibinfo {author} {\bibfnamefont {Q.-Y.}\
  \bibnamefont {Wang}}, \bibinfo {author} {\bibfnamefont {Z.}~\bibnamefont
  {Li}}, \bibinfo {author} {\bibfnamefont {L.}~\bibnamefont {Wang}}, \bibinfo
  {author} {\bibfnamefont {L.}~\bibnamefont {Zhao}}, \bibinfo {author}
  {\bibfnamefont {S.}~\bibnamefont {He}},  \emph {et~al.},\ }\href {\doibase
  10.1038/ncomms1946} {\bibfield  {journal} {\bibinfo  {journal} {Nat.
  Commun.}\ }\textbf {\bibinfo {volume} {3}},\ \bibinfo {pages} {931} (\bibinfo
  {year} {2012})},\ \Eprint {http://arxiv.org/abs/1202.5849} {arXiv:1202.5849}
  \BibitemShut {NoStop}%
\bibitem [{\citenamefont {Reyren}\ \emph {et~al.}(2007)\citenamefont {Reyren},
  \citenamefont {Thiel}, \citenamefont {Caviglia}, \citenamefont {Kourkoutis},
  \citenamefont {Hammerl}, \citenamefont {Richter}, \citenamefont {Schneider},
  \citenamefont {Kopp}, \citenamefont {R{\"u}etschi}, \citenamefont {Jaccard}
  \emph {et~al.}}]{triscone2007science}%
  \BibitemOpen
  \bibfield  {author} {\bibinfo {author} {\bibfnamefont {N.}~\bibnamefont
  {Reyren}}, \bibinfo {author} {\bibfnamefont {S.}~\bibnamefont {Thiel}},
  \bibinfo {author} {\bibfnamefont {A.}~\bibnamefont {Caviglia}}, \bibinfo
  {author} {\bibfnamefont {L.~F.}\ \bibnamefont {Kourkoutis}}, \bibinfo
  {author} {\bibfnamefont {G.}~\bibnamefont {Hammerl}}, \bibinfo {author}
  {\bibfnamefont {C.}~\bibnamefont {Richter}}, \bibinfo {author} {\bibfnamefont
  {C.}~\bibnamefont {Schneider}}, \bibinfo {author} {\bibfnamefont
  {T.}~\bibnamefont {Kopp}}, \bibinfo {author} {\bibfnamefont {A.-S.}\
  \bibnamefont {R{\"u}etschi}}, \bibinfo {author} {\bibfnamefont
  {D.}~\bibnamefont {Jaccard}},  \emph {et~al.},\ }\href {\doibase
  10.1126/science.1146006} {\bibfield  {journal} {\bibinfo  {journal}
  {Science}\ }\textbf {\bibinfo {volume} {317}},\ \bibinfo {pages} {1196}
  (\bibinfo {year} {2007})}\BibitemShut {NoStop}%
\bibitem [{\citenamefont {Nagamatsu}\ \emph {et~al.}(2001)\citenamefont
  {Nagamatsu}, \citenamefont {Nakagawa}, \citenamefont {Muranaka},
  \citenamefont {Zenitani},\ and\ \citenamefont {Akimitsu}}]{Nagamatsu2001}%
  \BibitemOpen
  \bibfield  {author} {\bibinfo {author} {\bibfnamefont {J.}~\bibnamefont
  {Nagamatsu}}, \bibinfo {author} {\bibfnamefont {N.}~\bibnamefont {Nakagawa}},
  \bibinfo {author} {\bibfnamefont {T.}~\bibnamefont {Muranaka}}, \bibinfo
  {author} {\bibfnamefont {Y.}~\bibnamefont {Zenitani}}, \ and\ \bibinfo
  {author} {\bibfnamefont {J.}~\bibnamefont {Akimitsu}},\ }\href {\doibase
  10.1038/35065039} {\bibfield  {journal} {\bibinfo  {journal} {Nature}\
  }\textbf {\bibinfo {volume} {410}},\ \bibinfo {pages} {63} (\bibinfo {year}
  {2001})}\BibitemShut {NoStop}%
\bibitem [{\citenamefont {Shimakage}\ \emph {et~al.}(2008)\citenamefont
  {Shimakage}, \citenamefont {Tatsumi},\ and\ \citenamefont
  {Wang}}]{Shimakage2008}%
  \BibitemOpen
  \bibfield  {author} {\bibinfo {author} {\bibfnamefont {H.}~\bibnamefont
  {Shimakage}}, \bibinfo {author} {\bibfnamefont {M.}~\bibnamefont {Tatsumi}},
  \ and\ \bibinfo {author} {\bibfnamefont {Z.}~\bibnamefont {Wang}},\ }\href
  {\doibase 10.1088/0953-2048/21/9/095009} {\bibfield  {journal} {\bibinfo
  {journal} {Supercond. Sci. Technol.}\ }\textbf {\bibinfo {volume} {21}},\
  \bibinfo {pages} {95009} (\bibinfo {year} {2008})}\BibitemShut {NoStop}%
\bibitem [{\citenamefont {Zhang}\ \emph
  {et~al.}(2010{\natexlab{b}})\citenamefont {Zhang}, \citenamefont {Lin},
  \citenamefont {Dai}, \citenamefont {Li}, \citenamefont {Wang}, \citenamefont
  {Zhang}, \citenamefont {Wang},\ and\ \citenamefont {Feng}}]{Zhang2010}%
  \BibitemOpen
  \bibfield  {author} {\bibinfo {author} {\bibfnamefont {Y.}~\bibnamefont
  {Zhang}}, \bibinfo {author} {\bibfnamefont {Z.}~\bibnamefont {Lin}}, \bibinfo
  {author} {\bibfnamefont {Q.}~\bibnamefont {Dai}}, \bibinfo {author}
  {\bibfnamefont {D.}~\bibnamefont {Li}}, \bibinfo {author} {\bibfnamefont
  {Y.}~\bibnamefont {Wang}}, \bibinfo {author} {\bibfnamefont {Y.}~\bibnamefont
  {Zhang}}, \bibinfo {author} {\bibfnamefont {Y.}~\bibnamefont {Wang}}, \ and\
  \bibinfo {author} {\bibfnamefont {Q.}~\bibnamefont {Feng}},\ }\href {\doibase
  10.1088/0953-2048/24/1/015013} {\bibfield  {journal} {\bibinfo  {journal}
  {Supercond. Sci. Technol.}\ }\textbf {\bibinfo {volume} {24}},\ \bibinfo
  {pages} {015013} (\bibinfo {year} {2010}{\natexlab{b}})},\ \Eprint
  {http://arxiv.org/abs/1102.5625} {arXiv:1102.5625} \BibitemShut {NoStop}%
\bibitem [{\citenamefont {Zhang}\ \emph {et~al.}(2013)\citenamefont {Zhang},
  \citenamefont {Wang}, \citenamefont {Wang}, \citenamefont {Zhang},
  \citenamefont {Liu}, \citenamefont {Feng},\ and\ \citenamefont
  {Gan}}]{Zhang2013}%
  \BibitemOpen
  \bibfield  {author} {\bibinfo {author} {\bibfnamefont {C.}~\bibnamefont
  {Zhang}}, \bibinfo {author} {\bibfnamefont {Y.}~\bibnamefont {Wang}},
  \bibinfo {author} {\bibfnamefont {D.}~\bibnamefont {Wang}}, \bibinfo {author}
  {\bibfnamefont {Y.}~\bibnamefont {Zhang}}, \bibinfo {author} {\bibfnamefont
  {Z.-H.}\ \bibnamefont {Liu}}, \bibinfo {author} {\bibfnamefont {Q.-R.}\
  \bibnamefont {Feng}}, \ and\ \bibinfo {author} {\bibfnamefont {Z.-Z.}\
  \bibnamefont {Gan}},\ }\href {\doibase 10.1063/1.4812738} {\bibfield
  {journal} {\bibinfo  {journal} {J. App. Phys.}\ }\textbf {\bibinfo {volume}
  {114}},\ \bibinfo {pages} {023903} (\bibinfo {year} {2013})}\BibitemShut
  {NoStop}%
\bibitem [{\citenamefont {Sza{\l}owski}(2006)}]{Karol2006}%
  \BibitemOpen
  \bibfield  {author} {\bibinfo {author} {\bibfnamefont {K.}~\bibnamefont
  {Sza{\l}owski}},\ }\href {\doibase 10.1103/PhysRevB.74.094501} {\bibfield
  {journal} {\bibinfo  {journal} {Phys. Rev. B}\ }\textbf {\bibinfo {volume}
  {74}},\ \bibinfo {pages} {094501} (\bibinfo {year} {2006})},\ \Eprint
  {http://arxiv.org/abs/1407.3717} {arXiv:1407.3717} \BibitemShut {NoStop}%
\bibitem [{\citenamefont {Bussmann-Holder}\ and\ \citenamefont
  {Bianconi}(2003)}]{bianconi2003}%
  \BibitemOpen
  \bibfield  {author} {\bibinfo {author} {\bibfnamefont {A.}~\bibnamefont
  {Bussmann-Holder}}\ and\ \bibinfo {author} {\bibfnamefont {A.}~\bibnamefont
  {Bianconi}},\ }\href {\doibase 10.1103/PhysRevB.67.132509} {\bibfield
  {journal} {\bibinfo  {journal} {Phys. Rev. B}\ }\textbf {\bibinfo {volume}
  {67}},\ \bibinfo {pages} {132509} (\bibinfo {year} {2003})}\BibitemShut
  {NoStop}%
\bibitem [{\citenamefont {Bianconi}\ \emph {et~al.}(2004)\citenamefont
  {Bianconi}, \citenamefont {Agrestini},\ and\ \citenamefont
  {Bussmann-Holder}}]{bianconi2004}%
  \BibitemOpen
  \bibfield  {author} {\bibinfo {author} {\bibfnamefont {A.}~\bibnamefont
  {Bianconi}}, \bibinfo {author} {\bibfnamefont {S.}~\bibnamefont {Agrestini}},
  \ and\ \bibinfo {author} {\bibfnamefont {A.}~\bibnamefont
  {Bussmann-Holder}},\ }\href {\doibase 10.1023/B:JOSC.0000021214.52321.ab}
  {\bibfield  {journal} {\bibinfo  {journal} {J. Supercond.}\ }\textbf
  {\bibinfo {volume} {17}},\ \bibinfo {pages} {205} (\bibinfo {year}
  {2004})}\BibitemShut {NoStop}%
\bibitem [{\citenamefont {Innocenti}\ \emph
  {et~al.}(2010{\natexlab{a}})\citenamefont {Innocenti}, \citenamefont
  {Poccia}, \citenamefont {Ricci}, \citenamefont {Valletta}, \citenamefont
  {Caprara}, \citenamefont {Perali},\ and\ \citenamefont
  {Bianconi}}]{bianconi2010}%
  \BibitemOpen
  \bibfield  {author} {\bibinfo {author} {\bibfnamefont {D.}~\bibnamefont
  {Innocenti}}, \bibinfo {author} {\bibfnamefont {N.}~\bibnamefont {Poccia}},
  \bibinfo {author} {\bibfnamefont {A.}~\bibnamefont {Ricci}}, \bibinfo
  {author} {\bibfnamefont {A.}~\bibnamefont {Valletta}}, \bibinfo {author}
  {\bibfnamefont {S.}~\bibnamefont {Caprara}}, \bibinfo {author} {\bibfnamefont
  {A.}~\bibnamefont {Perali}}, \ and\ \bibinfo {author} {\bibfnamefont
  {A.}~\bibnamefont {Bianconi}},\ }\href {\doibase 10.1103/PhysRevB.82.184528}
  {\bibfield  {journal} {\bibinfo  {journal} {Phys. Rev. B}\ }\textbf {\bibinfo
  {volume} {82}},\ \bibinfo {pages} {184528} (\bibinfo {year}
  {2010}{\natexlab{a}})},\ \Eprint {http://arxiv.org/abs/1007.0510}
  {arXiv:1007.0510} \BibitemShut {NoStop}%
\bibitem [{\citenamefont {Innocenti}\ \emph
  {et~al.}(2010{\natexlab{b}})\citenamefont {Innocenti}, \citenamefont
  {Caprara}, \citenamefont {Poccia}, \citenamefont {Ricci}, \citenamefont
  {Valletta},\ and\ \citenamefont {Bianconi}}]{bianconi2010a}%
  \BibitemOpen
  \bibfield  {author} {\bibinfo {author} {\bibfnamefont {D.}~\bibnamefont
  {Innocenti}}, \bibinfo {author} {\bibfnamefont {S.}~\bibnamefont {Caprara}},
  \bibinfo {author} {\bibfnamefont {N.}~\bibnamefont {Poccia}}, \bibinfo
  {author} {\bibfnamefont {A.}~\bibnamefont {Ricci}}, \bibinfo {author}
  {\bibfnamefont {A.}~\bibnamefont {Valletta}}, \ and\ \bibinfo {author}
  {\bibfnamefont {A.}~\bibnamefont {Bianconi}},\ }\href {\doibase
  10.1088/0953-2048/24/1/015012} {\bibfield  {journal} {\bibinfo  {journal}
  {Supercond. Sci. Technol.}\ }\textbf {\bibinfo {volume} {24}},\ \bibinfo
  {pages} {015012} (\bibinfo {year} {2010}{\natexlab{b}})},\ \Eprint
  {http://arxiv.org/abs/1011.4548} {arXiv:1011.4548} \BibitemShut {NoStop}%
\bibitem [{\citenamefont {Ara{\'u}jo}\ \emph {et~al.}(2011)\citenamefont
  {Ara{\'u}jo}, \citenamefont {Garc{\'\i}a-Garc{\'\i}a},\ and\ \citenamefont
  {Sacramento}}]{aggsacramento2011}%
  \BibitemOpen
  \bibfield  {author} {\bibinfo {author} {\bibfnamefont {M.~A.~N.}\
  \bibnamefont {Ara{\'u}jo}}, \bibinfo {author} {\bibfnamefont {A.~M.}\
  \bibnamefont {Garc{\'\i}a-Garc{\'\i}a}}, \ and\ \bibinfo {author}
  {\bibfnamefont {P.}~\bibnamefont {Sacramento}},\ }\href {\doibase
  10.1103/PhysRevB.84.172502} {\bibfield  {journal} {\bibinfo  {journal} {Phys.
  Rev. B}\ }\textbf {\bibinfo {volume} {84}},\ \bibinfo {pages} {172502}
  (\bibinfo {year} {2011})},\ \Eprint {http://arxiv.org/abs/1103.3290}
  {arXiv:1103.3290} \BibitemShut {NoStop}%
\bibitem [{\citenamefont {Brack}\ and\ \citenamefont {Bhaduri}(1997)}]{Brack}%
  \BibitemOpen
  \bibfield  {author} {\bibinfo {author} {\bibfnamefont {M.}~\bibnamefont
  {Brack}}\ and\ \bibinfo {author} {\bibfnamefont {R.}~\bibnamefont
  {Bhaduri}},\ }\href@noop {} {\emph {\bibinfo {title} {Semicalssical
  Physics}}}\ (\bibinfo  {publisher} {Addison-Wesley},\ \bibinfo {year}
  {1997})\BibitemShut {NoStop}%
\bibitem [{\citenamefont {Apostol}(1973)}]{Apostol}%
  \BibitemOpen
  \bibfield  {author} {\bibinfo {author} {\bibfnamefont {T.~M.}\ \bibnamefont
  {Apostol}},\ }\href@noop {} {\emph {\bibinfo {title} {Mathematical
  Analysis}}}\ (\bibinfo  {publisher} {Addison-Wesley},\ \bibinfo {year}
  {1973})\BibitemShut {NoStop}%
\bibitem [{\citenamefont {BenDaniel}\ and\ \citenamefont
  {Duke}(1966)}]{BenDaniel1966}%
  \BibitemOpen
  \bibfield  {author} {\bibinfo {author} {\bibfnamefont {D.}~\bibnamefont
  {BenDaniel}}\ and\ \bibinfo {author} {\bibfnamefont {C.}~\bibnamefont
  {Duke}},\ }\href {\doibase 10.1103/PhysRev.152.683} {\bibfield  {journal}
  {\bibinfo  {journal} {Phys. Rev.}\ }\textbf {\bibinfo {volume} {152}},\
  \bibinfo {pages} {683} (\bibinfo {year} {1966})}\BibitemShut {NoStop}%
\bibitem [{\citenamefont {Garc{\'\i}a-Garc{\'\i}a}\ \emph
  {et~al.}(2011)\citenamefont {Garc{\'\i}a-Garc{\'\i}a}, \citenamefont
  {Urbina}, \citenamefont {Yuzbashyan}, \citenamefont {Richter},\ and\
  \citenamefont {Altshuler}}]{Garcia2011}%
  \BibitemOpen
  \bibfield  {author} {\bibinfo {author} {\bibfnamefont {A.}~\bibnamefont
  {Garc{\'\i}a-Garc{\'\i}a}}, \bibinfo {author} {\bibfnamefont
  {J.}~\bibnamefont {Urbina}}, \bibinfo {author} {\bibfnamefont
  {E.}~\bibnamefont {Yuzbashyan}}, \bibinfo {author} {\bibfnamefont
  {K.}~\bibnamefont {Richter}}, \ and\ \bibinfo {author} {\bibfnamefont
  {B.}~\bibnamefont {Altshuler}},\ }\href {\doibase 10.1103/PhysRevB.83.014510}
  {\bibfield  {journal} {\bibinfo  {journal} {Phys. Rev. B}\ }\textbf {\bibinfo
  {volume} {83}},\ \bibinfo {pages} {014510} (\bibinfo {year} {2011})},\
  \Eprint {http://arxiv.org/abs/0911.1559} {arXiv:0911.1559} \BibitemShut
  {NoStop}%
\bibitem [{\citenamefont {Eom}\ \emph {et~al.}(2006)\citenamefont {Eom},
  \citenamefont {Qin}, \citenamefont {Chou},\ and\ \citenamefont
  {Shih}}]{Eom2006}%
  \BibitemOpen
  \bibfield  {author} {\bibinfo {author} {\bibfnamefont {D.}~\bibnamefont
  {Eom}}, \bibinfo {author} {\bibfnamefont {S.}~\bibnamefont {Qin}}, \bibinfo
  {author} {\bibfnamefont {M.}~\bibnamefont {Chou}}, \ and\ \bibinfo {author}
  {\bibfnamefont {C.~K.}\ \bibnamefont {Shih}},\ }\href {\doibase
  10.1103/PhysRevLett.96.027005} {\bibfield  {journal} {\bibinfo  {journal}
  {Phys. Rev. Lett.}\ }\textbf {\bibinfo {volume} {96}},\ \bibinfo {pages}
  {027005} (\bibinfo {year} {2006})}\BibitemShut {NoStop}%
\bibitem [{\citenamefont {Matveev}\ and\ \citenamefont
  {Larkin}(1997)}]{matveev1997}%
  \BibitemOpen
  \bibfield  {author} {\bibinfo {author} {\bibfnamefont {K.}~\bibnamefont
  {Matveev}}\ and\ \bibinfo {author} {\bibfnamefont {A.}~\bibnamefont
  {Larkin}},\ }\href {\doibase 10.1103/PhysRevLett.78.3749} {\bibfield
  {journal} {\bibinfo  {journal} {Phys. Rev. Lett.}\ }\textbf {\bibinfo
  {volume} {78}},\ \bibinfo {pages} {3749} (\bibinfo {year} {1997})},\ \Eprint
  {http://arxiv.org/abs/9701041} {cond-mat:9701041} \BibitemShut {NoStop}%
\bibitem [{\citenamefont {Yuzbashyan}\ \emph {et~al.}(2005)\citenamefont
  {Yuzbashyan}, \citenamefont {Baytin},\ and\ \citenamefont
  {Altshuler}}]{Yuzbashyan2005}%
  \BibitemOpen
  \bibfield  {author} {\bibinfo {author} {\bibfnamefont {E.~A.}\ \bibnamefont
  {Yuzbashyan}}, \bibinfo {author} {\bibfnamefont {A.~A.}\ \bibnamefont
  {Baytin}}, \ and\ \bibinfo {author} {\bibfnamefont {B.~L.}\ \bibnamefont
  {Altshuler}},\ }\href {\doibase 10.1103/PhysRevB.71.094505} {\bibfield
  {journal} {\bibinfo  {journal} {Phys. Rev. B}\ }\textbf {\bibinfo {volume}
  {71}},\ \bibinfo {pages} {094505} (\bibinfo {year} {2005})}\BibitemShut
  {NoStop}%
\bibitem [{\citenamefont {M{\"u}hlschlegel}\ \emph {et~al.}(1972)\citenamefont
  {M{\"u}hlschlegel}, \citenamefont {Scalapino},\ and\ \citenamefont
  {Denton}}]{Muhlschlegel1972}%
  \BibitemOpen
  \bibfield  {author} {\bibinfo {author} {\bibfnamefont {B.}~\bibnamefont
  {M{\"u}hlschlegel}}, \bibinfo {author} {\bibfnamefont {D.}~\bibnamefont
  {Scalapino}}, \ and\ \bibinfo {author} {\bibfnamefont {R.}~\bibnamefont
  {Denton}},\ }\href {\doibase 10.1103/PhysRevB.6.1767} {\bibfield  {journal}
  {\bibinfo  {journal} {Phys. Rev. B}\ }\textbf {\bibinfo {volume} {6}},\
  \bibinfo {pages} {1767} (\bibinfo {year} {1972})}\BibitemShut {NoStop}%
\bibitem [{\citenamefont {Kong}\ \emph {et~al.}(2001)\citenamefont {Kong},
  \citenamefont {Dolgov}, \citenamefont {Jepsen},\ and\ \citenamefont
  {Andersen}}]{Kong2001}%
  \BibitemOpen
  \bibfield  {author} {\bibinfo {author} {\bibfnamefont {Y.}~\bibnamefont
  {Kong}}, \bibinfo {author} {\bibfnamefont {O.}~\bibnamefont {Dolgov}},
  \bibinfo {author} {\bibfnamefont {O.}~\bibnamefont {Jepsen}}, \ and\ \bibinfo
  {author} {\bibfnamefont {O.}~\bibnamefont {Andersen}},\ }\href {\doibase
  10.1103/PhysRevB.64.020501} {\bibfield  {journal} {\bibinfo  {journal} {Phys.
  Rev. B}\ }\textbf {\bibinfo {volume} {64}},\ \bibinfo {pages} {020501}
  (\bibinfo {year} {2001})},\ \Eprint {http://arxiv.org/abs/0102499}
  {cond-mat:0102499} \BibitemShut {NoStop}%
\bibitem [{\citenamefont {Kirchmann}\ \emph {et~al.}(2010)\citenamefont
  {Kirchmann}, \citenamefont {Rettig}, \citenamefont {Zubizarreta},
  \citenamefont {Silkin}, \citenamefont {Chulkov},\ and\ \citenamefont
  {Bovensiepen}}]{Kirchmann2010}%
  \BibitemOpen
  \bibfield  {author} {\bibinfo {author} {\bibfnamefont {P.~S.}\ \bibnamefont
  {Kirchmann}}, \bibinfo {author} {\bibfnamefont {L.}~\bibnamefont {Rettig}},
  \bibinfo {author} {\bibfnamefont {X.}~\bibnamefont {Zubizarreta}}, \bibinfo
  {author} {\bibfnamefont {V.~M.}\ \bibnamefont {Silkin}}, \bibinfo {author}
  {\bibfnamefont {E.~V.}\ \bibnamefont {Chulkov}}, \ and\ \bibinfo {author}
  {\bibfnamefont {U.}~\bibnamefont {Bovensiepen}},\ }\href {\doibase
  10.1038/nphys1735} {\bibfield  {journal} {\bibinfo  {journal} {Nat. Phys.}\
  }\textbf {\bibinfo {volume} {6}},\ \bibinfo {pages} {782} (\bibinfo {year}
  {2010})}\BibitemShut {NoStop}%
\bibitem [{\citenamefont {Golubov}\ \emph {et~al.}(2002)\citenamefont
  {Golubov}, \citenamefont {Kortus}, \citenamefont {Dolgov}, \citenamefont
  {Jepsen},\ and\ \citenamefont {...}}]{Golubov2002}%
  \BibitemOpen
  \bibfield  {author} {\bibinfo {author} {\bibfnamefont {A.}~\bibnamefont
  {Golubov}}, \bibinfo {author} {\bibfnamefont {J.}~\bibnamefont {Kortus}},
  \bibinfo {author} {\bibfnamefont {O.}~\bibnamefont {Dolgov}}, \bibinfo
  {author} {\bibfnamefont {O.}~\bibnamefont {Jepsen}}, \ and\ \bibinfo {author}
  {\bibnamefont {...}},\ }\href {\doibase 10.1088/0953-8984/14/6/320}
  {\bibfield  {journal} {\bibinfo  {journal} {J. Phys. Condens. Matter}\
  }\textbf {\bibinfo {volume} {14}},\ \bibinfo {pages} {1353} (\bibinfo {year}
  {2002})}\BibitemShut {NoStop}%
\bibitem [{\citenamefont {Liu}\ \emph {et~al.}(2001)\citenamefont {Liu},
  \citenamefont {Mazin},\ and\ \citenamefont {Kortus}}]{Liu2001}%
  \BibitemOpen
  \bibfield  {author} {\bibinfo {author} {\bibfnamefont {A.}~\bibnamefont
  {Liu}}, \bibinfo {author} {\bibfnamefont {I.}~\bibnamefont {Mazin}}, \ and\
  \bibinfo {author} {\bibfnamefont {J.}~\bibnamefont {Kortus}},\ }\href
  {\doibase 10.1103/PhysRevLett.87.087005} {\bibfield  {journal} {\bibinfo
  {journal} {Phys. Rev. Lett.}\ }\textbf {\bibinfo {volume} {87}},\ \bibinfo
  {pages} {087005} (\bibinfo {year} {2001})},\ \Eprint
  {http://arxiv.org/abs/0103570} {cond-mat:0103570} \BibitemShut {NoStop}%
\bibitem [{\citenamefont {Parmenter}(1968)}]{parmenter}%
  \BibitemOpen
  \bibfield  {author} {\bibinfo {author} {\bibfnamefont {R.}~\bibnamefont
  {Parmenter}},\ }\href {\doibase 10.1103/PhysRev.166.392} {\bibfield
  {journal} {\bibinfo  {journal} {Phys. Rev.}\ }\textbf {\bibinfo {volume}
  {166}},\ \bibinfo {pages} {392} (\bibinfo {year} {1968})}\BibitemShut
  {NoStop}%
\bibitem [{\citenamefont {Garc{\'\i}a-Garc{\'\i}a}\ \emph
  {et~al.}(2008)\citenamefont {Garc{\'\i}a-Garc{\'\i}a}, \citenamefont
  {Urbina}, \citenamefont {Yuzbashyan}, \citenamefont {Richter},\ and\
  \citenamefont {Altshuler}}]{aggalt2008}%
  \BibitemOpen
  \bibfield  {author} {\bibinfo {author} {\bibfnamefont {A.~M.}\ \bibnamefont
  {Garc{\'\i}a-Garc{\'\i}a}}, \bibinfo {author} {\bibfnamefont {J.~D.}\
  \bibnamefont {Urbina}}, \bibinfo {author} {\bibfnamefont {E.~A.}\
  \bibnamefont {Yuzbashyan}}, \bibinfo {author} {\bibfnamefont
  {K.}~\bibnamefont {Richter}}, \ and\ \bibinfo {author} {\bibfnamefont
  {B.~L.}\ \bibnamefont {Altshuler}},\ }\href {\doibase
  10.1103/PhysRevLett.100.187001} {\bibfield  {journal} {\bibinfo  {journal}
  {Phys. Rev. Lett.}\ }\textbf {\bibinfo {volume} {100}},\ \bibinfo {pages}
  {187001} (\bibinfo {year} {2008})}\BibitemShut {NoStop}%
\bibitem [{\citenamefont {Moshchalkov}\ \emph {et~al.}(2009)\citenamefont
  {Moshchalkov}, \citenamefont {Menghini}, \citenamefont {Nishio},
  \citenamefont {Chen}, \citenamefont {Silhanek}, \citenamefont {Dao},
  \citenamefont {Chibotaru}, \citenamefont {Zhigadlo},\ and\ \citenamefont
  {Karpinski}}]{Moshchalkov2009}%
  \BibitemOpen
  \bibfield  {author} {\bibinfo {author} {\bibfnamefont {V.}~\bibnamefont
  {Moshchalkov}}, \bibinfo {author} {\bibfnamefont {M.}~\bibnamefont
  {Menghini}}, \bibinfo {author} {\bibfnamefont {T.}~\bibnamefont {Nishio}},
  \bibinfo {author} {\bibfnamefont {Q.}~\bibnamefont {Chen}}, \bibinfo {author}
  {\bibfnamefont {A.}~\bibnamefont {Silhanek}}, \bibinfo {author}
  {\bibfnamefont {V.}~\bibnamefont {Dao}}, \bibinfo {author} {\bibfnamefont
  {L.}~\bibnamefont {Chibotaru}}, \bibinfo {author} {\bibfnamefont
  {N.}~\bibnamefont {Zhigadlo}}, \ and\ \bibinfo {author} {\bibfnamefont
  {J.}~\bibnamefont {Karpinski}},\ }\href {\doibase
  10.1103/PhysRevLett.102.117001} {\bibfield  {journal} {\bibinfo  {journal}
  {Phys. Rev. Lett.}\ }\textbf {\bibinfo {volume} {102}},\ \bibinfo {pages}
  {117001} (\bibinfo {year} {2009})},\ \Eprint {http://arxiv.org/abs/0902.0997}
  {arXiv:0902.0997} \BibitemShut {NoStop}%
\bibitem [{\citenamefont {Brinkman}\ \emph {et~al.}(2002)\citenamefont
  {Brinkman}, \citenamefont {Gloubov}, \citenamefont {Rogalla}, \citenamefont
  {Dolgov}, \citenamefont {Kortus}, \citenamefont {Kong}, \citenamefont
  {Jepsen},\ and\ \citenamefont {Andersen}}]{Brinkman2002}%
  \BibitemOpen
  \bibfield  {author} {\bibinfo {author} {\bibfnamefont {A.}~\bibnamefont
  {Brinkman}}, \bibinfo {author} {\bibfnamefont {A.~A.}\ \bibnamefont
  {Gloubov}}, \bibinfo {author} {\bibfnamefont {H.}~\bibnamefont {Rogalla}},
  \bibinfo {author} {\bibfnamefont {O.}~\bibnamefont {Dolgov}}, \bibinfo
  {author} {\bibfnamefont {J.}~\bibnamefont {Kortus}}, \bibinfo {author}
  {\bibfnamefont {Y.}~\bibnamefont {Kong}}, \bibinfo {author} {\bibfnamefont
  {O.}~\bibnamefont {Jepsen}}, \ and\ \bibinfo {author} {\bibfnamefont
  {O.}~\bibnamefont {Andersen}},\ }\href {\doibase 10.1103/PhysRevB.65.180517}
  {\bibfield  {journal} {\bibinfo  {journal} {Phys. Rev. B}\ }\textbf {\bibinfo
  {volume} {65}},\ \bibinfo {pages} {180517} (\bibinfo {year} {2002})},\
  \Eprint {http://arxiv.org/abs/0111115} {cond-mat:0111115} \BibitemShut
  {NoStop}%
\bibitem [{\citenamefont {Ozer}\ \emph {et~al.}(2006)\citenamefont {Ozer},
  \citenamefont {Thompson},\ and\ \citenamefont {Weitering}}]{Ozer2006}%
  \BibitemOpen
  \bibfield  {author} {\bibinfo {author} {\bibfnamefont {M.~M.}\ \bibnamefont
  {Ozer}}, \bibinfo {author} {\bibfnamefont {J.~R.}\ \bibnamefont {Thompson}},
  \ and\ \bibinfo {author} {\bibfnamefont {H.~H.}\ \bibnamefont {Weitering}},\
  }\href {\doibase 10.1038/nphys244} {\bibfield  {journal} {\bibinfo  {journal}
  {Nat. Phys.}\ }\textbf {\bibinfo {volume} {2}},\ \bibinfo {pages} {173}
  (\bibinfo {year} {2006})},\ \Eprint {http://arxiv.org/abs/0601641}
  {cond-mat:0601641} \BibitemShut {NoStop}%
\end{thebibliography}%

\end{document}